\documentclass[english,prd,onecolumn,superscriptaddress,nofootinbib,preprintnumbers]{revtex4}
\usepackage[latin1]{inputenc}
\usepackage{graphicx}
\usepackage{amsmath,amsthm,amssymb}

\makeatletter

\usepackage{amsfonts}
\usepackage{dcolumn}
\usepackage{hyperref}

\def\be{\begin{equation}}
\def\ee{\end{equation}}
\def\ba{\begin{eqnarray}}
\def\ea{\end{eqnarray}}
\def\bs{\begin{subequations}}
\def\es{\end{subequations}}

\newcommand{\rd}{{\rm d}}

\usepackage{color}

\makeatother

\usepackage{babel}
\makeatother
\begin{document}

\title{Chameleon scalar fields in relativistic gravitational backgrounds}

\author{Shinji Tsujikawa}
\affiliation{Department of Physics, Faculty of Science, Tokyo University of Science,
1-3, Kagurazaka, Shinjuku-ku, Tokyo 162-8601, Japan}
\email{shinji@rs.kagu.tus.ac.jp}

\author{Takashi Tamaki}
\address{Department of Physics, Waseda University, Okubo 3-4-1, Tokyo 
169-8555, Japan}
\email{tamaki@gravity.phys.waseda.ac.jp}
\affiliation{Department of Physics, Rikkyo University, Toshima, Tokyo 171-8501, Japan}

\author{Reza Tavakol}
\affiliation{Astronomy Unit, School of Mathematical Sciences,
Queen Mary University of London,
London E1 4NS, UK}
\email{r.tavakol@qmul.ac.uk}

\begin{abstract}

We study the field profile of a scalar field $\phi$ that couples to a matter fluid
(dubbed a chameleon field) in the relativistic gravitational
background of a spherically symmetric spacetime.
Employing a linear expansion in terms of the gravitational potential $\Phi_c$ 
at the surface of a compact object with a constant density, 
we derive the thin-shell field profile both inside and outside the object, 
as well as the resulting effective coupling with matter, analytically.
We also carry out numerical simulations for the class of
inverse power-law potentials $V(\phi)=M^{4+n} \phi^{-n}$ 
by employing the information provided
by our analytical solutions to set the
boundary conditions around the centre of the object
and show that thin-shell solutions in fact exist if the gravitational potential 
$\Phi_c$ is smaller than 0.3, 
which marginally covers the case of neutron stars.
Thus the chameleon mechanism is present in the relativistic 
gravitational backgrounds, capable of reducing the effective coupling.
Since thin-shell solutions are sensitive to the choice of boundary conditions, 
our analytic field profile is very helpful to provide appropriate 
boundary conditions for $\Phi_c \lesssim O(0.1)$.

\end{abstract}

\date{\today}

\maketitle

\section{Introduction}

The origin of the so called dark energy responsible for the present 
cosmic acceleration remains a great mystery.
Since the cosmological constant (originating from the vacuum energy)
is plagued by a severe fine-tuning problem, many alternative models
have been proposed to account for the origin of dark energy 
(see Refs.~\cite{review} for reviews).
A number of these models, including the quintessence \cite{quin}, 
k-essence \cite{kes} and tachyon \cite{tac} models, make use 
of a scalar field with a very light mass ($m_\phi \sim 10^{-33}$~eV)
in order to account for the present cosmic acceleration.
If the scalar field originates from candidate theories for
fundamental interactions such as string theory or supergravity, 
it should interact with the standard model 
particles with a long ranged force (the so called ``fifth force'').
In string theory, for example, a dilaton field universally couples
to matter as well as gravity \cite{Lidsey}.
Similarly, in modified gravity theories such as $f(R)$ gravity \cite{fR} 
and scalar-tensor theories \cite{stensor}, the scalar degree of freedom 
interacts with the matter fluid (except for radiation). 
This is clearly seen if one transforms 
the action to the Einstein frame via a conformal
transformation \cite{Maeda}. For example, it is known that 
Brans-Dicke theory \cite{Brans} (a historically important
class of scalar-tensor theories) gives rise to a constant coupling 
$Q$ between the scalar field and 
the matter \cite{TUMTY}.
In this sense such modified gravity theories can be regarded as
a coupled quintessence scenario \cite{Amen} in the Einstein frame.

In the absence of a scalar-field potential, the present solar-system tests
constrain the strength of the coupling $Q$ to be smaller than 
the order of $10^{-3}$ \cite{TUMTY}.
However, the couplings that appear in string theory \cite{Lidsey}
and $f(R)$ gravity \cite{APT} are typically of the order of unity. 
In such cases it is not possible to satisfy the local gravity constraints,
unless a scalar-field potential with a large mass exists 
to suppress the coupling in the regions of high density.
Moreover, if the same field is responsible for the cosmic 
acceleration today, the potential needs to be sufficiently 
flat in the regions of low density (i.e., on cosmological scales).

In spite of the above requirements it is possible for the
large coupling models to satisfy the local gravity constraints through 
the chameleon mechanism \cite{KW,KW2}, while at the same time for the
field to have sufficiently small mass to lead to the present cosmic 
acceleration. The existence of a matter coupling gives rise to an extremum 
of the scalar-field potential around which the field can be 
stabilized. In high density regions, such as the interiors of the
astrophysical objects, the field mass about the extremum would be 
sufficiently large to avoid the propagation of the fifth force.
Meanwhile, the field would have a much lighter mass in the low-density 
environments, far away from compact objects, so that 
it could be responsible for the present cosmic acceleration.
In the case of inverse power-law potentials $V(\phi)=M^{4+n} 
\phi^{-n}$ \cite{Ratra} with $n \ge 1$, local gravity constraints can be
satisfied for $M \lesssim 10^{-2}$~eV \cite{KW2}.
Interestingly, this roughly corresponds to the energy scale
required for the cosmic acceleration today.
See Refs.~\cite{Brax,Gubser,fRca,Clifton,Das,radion,Mota,exper,TT}
for works concerning a number of interesting aspects of the 
chameleon mechanism.

So far the analyses of the chameleon mechanism have typically 
concentrated on the weak gravity backgrounds where the spherically 
symmetric metric is described by a Minkowski spacetime.
This amounts to neglecting the backreaction of gravitational potential
 on the scalar-field equation. 
In Ref.~\cite{TT} the field profile in the Minkowski background 
was analytically derived both inside and outside the object 
by taking into account the mass of the chameleon field inside the body.
In this settings it has been shown that the field would need to 
be extremely close to the maximum of the effective potential around the centre
of the spherically symmetric body in order to allow thin-shell solutions
required for consistency with the local gravity constraints. 

If we take into account the backreaction of gravitational potential
to the field equation, the relativistic pressure is present even 
in weak gravity backgrounds such as the Sun or the Earth.
It is expected that this effect changes the field profile
inside the body in order to allow the existence of thin-shell solutions.
We shall analytically derive the thin-shell field profile using a linear
expansion in terms of the gravitational potential $\Phi_c~(\ll 1)$ 
at the surface of compact objects.
In fact we show that there exists a region around the centre of 
the massive objects in which the field evolves toward the maximum 
of the effective potential because of the presence of the relativistic 
pressure. In order to realize thin-shell solutions, the driving force
along the potential needs to dominate over the pressure 
for distances larger than a critical value $ r= r_3$.
This distance ($r_3$) is required to be smaller than the distance $r_1$
at which the field enters a thin-shell regime.
In spite of such different properties of the field profile inside the body 
relative to the case of the Minkowski background, 
the effective coupling $Q_{\rm eff}$ outside the body can be reduced
by the presence of thin-shell solutions.
We confirm this by using numerical simulations for a class of potentials 
of the form $V(\phi)=M^{4+n}\phi^{-n}$.

To study the viability of theories with large couplings, 
it is important to determine whether thin-shell solutions can also exist 
in strong gravitational backgrounds with $\Phi_c \lesssim {\cal O}(0.1)$.
We shall derive analytic solutions using linear expansions in terms
of $\Phi_c$ and then carry out numerical simulations 
to confirm the validity of solutions in the regimes with 
$\Phi_c \lesssim O(0.1)$. Our analytic solutions are useful as a way of 
finding the boundary conditions around the centre of the object 
in order to obtain thin-shell solutions.
By choosing boundary conditions with field values larger than those
estimated by the analytic solutions, we shall demonstrate numerically
that the thin-shell solutions are present for backgrounds with 
gravitational potentials satisfying $\Phi_c \lesssim 0.3$, 
in the case of the field potentials of the type $V(\phi)=M^{4+n}\phi^{-n}$.
This marginally covers the case of neutron stars. 
In backgrounds with still larger gravitational potentials
the relativistic pressure around the centre 
of the object is so strong that the field typically overshoots the maximum of 
the effective potential to reach the singularity at $\phi=0$, 
unless the boundary conditions of the field around the centre of 
the body are chosen to be far from the maximum of the effective potential.
This overshoot behaviour is similar to the one recently found by Kobayashi and 
Maeda \cite{Kobayashi} in the context of $f(R)$ dark energy models
(see also Ref.~\cite{Frolov}).
We note, however, that our analytic solutions based on the linear expansion 
of $\Phi_c$ do not cover the field profiles for the really strong gravitational 
backgrounds with $\Phi_c=O(1)$. In such cases we need a separate analysis
which incorporates the formation of black holes.

The outline of the paper is as follows. In Section \ref{setup} we 
discuss our theoretical set up as well as giving the relevant equations
for the case of a spherically symmetric central body.
In section \ref{chame} we give the analytical thin-shell 
solutions to the scalar field equations, 
both inside and outside of the body, and
consider in turn the matching of thin-shell solutions.
In Section \ref{profile} we study the analytical field 
profile in more details and discuss how the field 
evolves as a function of $r$ in the presence of
the relativistic pressure.
In Section \ref{numerics} we integrate the field equation 
numerically and show the existence of thin-shell solutions
for $\Phi_c \lesssim 0.3$.
Finally Section \ref{conclude}
contains our conclusions.

\section{Setup}
\label{setup}

We consider settings in which a scalar field $\phi$ with 
potential $V(\phi)$ couples to a matter with a Lagrangian density ${\cal L}_m$. 
In particular we shall study theories based on the action
\begin{eqnarray}
\label{action}
S &=& \int {\rm d}^4 x\sqrt{-g} 
\left[ \frac{M_{\rm pl}^2}{2} R
-\frac12 (\nabla \phi)^2-V(\phi) \right]
-\int {\rm d}^4x\,{\cal L}_m 
(\Psi_m^{(i)},g_{\mu \nu}^{(i)})\,,
\end{eqnarray}
where $g$ is the determinant of the metric $g_{\mu \nu}$, 
$M_{\rm pl}=1/\sqrt{8 \pi G}$ is the reduced Planck mass
($G$ is the gravitational constant), 
$R$ is a Ricci scalar, and $\Psi_m^{(i)}$ are matter fields
that couple to a metric $g_{\mu \nu}^{(i)}$ related with 
the Einstein frame metric $g_{\mu \nu}$ via
\begin{eqnarray}
\label{gmunu1}
g_{\mu \nu}^{(i)}=e^{2 Q_i \phi} g_{\mu \nu}\,.
\end{eqnarray}
Here $Q_i$ are the strength of couplings for each matter field. 
In the following we shall consider cases in which the couplings 
are the same for each matter component, i.e., $Q_i=Q$, and use 
units such that $M_{\rm pl}=1/\sqrt{8\pi G}=1$. 
We restore $G$ when it is needed.

An example of a scalar-tensor theory which gives rise 
to constant couplings $Q$ in the Einstein frame is given by
the action \cite{TUMTY}
\begin{eqnarray}
\label{actionsca}
\tilde{S} = \int {\rm d}^4 x\sqrt{-\tilde{g}} 
\biggl[ \frac12 e^{-2 Q \phi} \tilde{R}
-\frac12 (1-6Q^2) e^{-2 Q \phi} (\tilde{\nabla} \phi)^2
-U(\phi) \biggr] -\int {\rm d}^4x\,{\cal L}_m 
(\Psi_m, \tilde{g}_{\mu \nu})\,,
\end{eqnarray}
where a tilde represents quantities in the Jordan frame.
The action (\ref{actionsca}) is equivalent to 
that in Brans-Dicke theory with a potential $U(\phi)$.
Under the conformal transformation, $g_{\mu \nu}=
e^{-2Q \phi} \tilde{g}_{\mu \nu}$, we obtain the 
action (\ref{action}) in the Einstein frame, together 
with the field potential $V(\phi)=U(\phi)\,e^{4Q \phi}$.
Clearly the metric $g_{\mu \nu}^{(i)}$ in Eq.~(\ref{gmunu1}) 
corresponds to the metric $\tilde{g}_{\mu \nu}$ in the Jordan frame.

To study chameleon fields in the relativistic gravitational background 
of a spherically symmetric body, we consider the following 
spherically symmetric static metric in the Einstein frame:
\begin{eqnarray}
\label{smetric}
\rd s^2=-e^{2\Psi (r)} \rd t^2+e^{2\Phi (r)} \rd r^2+
r^2 \rd \theta^2+r^2 \sin^2 \theta \rd \phi^2\,,
\end{eqnarray}
where $\Psi (r)$ and $\Phi (r)$ are functions of the
distance $r$ from the centre of symmetry.
For the action (\ref{action}) the energy 
momentum tensors for the scalar field $\phi$
and the matter are given, respectively, by 
\begin{eqnarray}
T_{\mu \nu}^{(\phi)} &=&
\partial_{\mu} \phi \partial_{\nu} \phi-g_{\mu \nu}
\left[ \frac12 g^{\alpha \beta} \partial_{\alpha} \phi
\partial_{\beta} \phi +V(\phi) \right]\,, \\
T_{\mu \nu}^{(m)} &=& \frac{2}{\sqrt{-g}}
\frac{\delta {\cal L}_m}{\delta g^{\mu \nu}}\,.
\end{eqnarray}
Under the gravitational background (\ref{smetric}),
the (00) and $(11)$ components for the energy 
momentum tensors are
\begin{eqnarray}
T^{0(\phi)}_0=-\frac12 e^{-2\Phi} \phi'^2
-V(\phi)\,,\quad
T^{1(\phi)}_1 = \frac12 e^{-2\Phi} \phi'^2
-V(\phi)\,,
\end{eqnarray}
where a prime represents a derivative with respect 
to $r$ and 
\begin{eqnarray}
T^{0(m)}_0 = e^{4Q \phi} \tilde{T}^{0(m)}_0\,,\quad
T^{1(m)}_1 = e^{4Q \phi} \tilde{T}^{r(m)}_r\,.
\end{eqnarray}
Here $\tilde{T}^{0(m)}_0$ and $\tilde{T}^{1(m)}_1$ are the 
energy momentum tensors of matter in the Jordan frame.
Denoting the energy density and the pressure of the matter
in the Jordan frame as $\tilde{\rho}_m$ and $\tilde{p}_m$, 
the matter energy-momentum tensor in this frame takes the form
$\tilde{T}^{\mu}_{\nu}=(-\tilde{\rho}_m,\tilde{p}_m,\tilde{p}_m,\tilde{p}_m)$.
The corresponding expressions for the energy density and pressure 
in the Einstein frame are then given by
$\rho_m=e^{4Q \phi} \tilde{\rho}_m$ and 
$p_m=e^{4Q \phi} \tilde{p}_m$.

The evolution equation for the scalar field $\phi$ is given by 
\begin{eqnarray}
\frac{\partial}{\partial x^i} \frac{\partial (\sqrt{-g}{\cal L}_{\phi})}
{\partial (\partial \phi/\partial x^i)}-\frac{\partial (\sqrt{-g}{\cal L}_\phi)}
{\partial \phi}-\frac{\partial {\cal L}_m}{\partial \phi}=0\,,
\end{eqnarray}
where the derivative of ${\cal L}_m={\cal L}_m (\tilde{g}_{\mu \nu})=
{\cal L}_m (e^{2Q\phi} g_{\mu \nu})$ in terms of $\phi$ is 
\begin{eqnarray}
\frac{\partial {\cal L}_m}{\partial \phi}
=\sqrt{-g}Q e^{4Q \phi} \tilde{g}_{\mu \nu} 
\tilde{T}^{\mu \nu}
=\sqrt{-g}Q (-\rho_m+3 p_m)\,.
\end{eqnarray}
We then obtain 
\begin{eqnarray}
\label{be1}
\phi''+\left( \frac{2}{r}+\Psi'-\Phi' \right) \phi'=
e^{2\Phi} \left[ V_{,\phi}+Q(\rho_m-3 p_m) \right]\,.
\end{eqnarray}
The Einstein equations give:
\begin{eqnarray}
\label{be2}
& &\Phi' = \frac{1-e^{2\Phi}}{2r}+4\pi Gr 
\left[ \frac12 \phi'^2+e^{2\Phi}V(\phi)+e^{2\Phi}\rho_m \right]\,, \\
\label{be3}
& &\Psi' = \frac{e^{2\Phi}-1}{2r}+4\pi Gr
\left[ \frac12 \phi'^2-e^{2\Phi}V(\phi)+e^{2\Phi}p_m \right]\,, \\
\label{be4}
& & \Psi''+\Psi'^2-\Psi' \Phi'+\frac{\Psi'-\Phi'}{r}
= -8\pi G \left[ \frac12 \phi'^2+
e^{2\Phi}V(\phi)-e^{2\Phi}p_m \right]\,.
\end{eqnarray}
{}From the conservation equation, 
$\nabla_{\mu} T^{\mu}_1=0$, we also obtain
\begin{eqnarray}
\label{be5}
p_m'+(\rho_m+p_m)\Psi'
+Q\phi' (\rho_m-3p_m)=0\,,
\end{eqnarray}
which is the generalization of the Tolman-Oppenheimer-Volkoff equation.
Note that this equation can also be derived by combining 
Eqs.~(\ref{be1})-(\ref{be4}).

Our main interest is the case in which the field potential $V(\phi)$
is responsible for dark energy.
In that case both $V(\phi)$ and $\phi'^2$ are negligible 
relative to $\rho_m$ in the local regions
whose density is much larger than the cosmological one
($\rho_0 \sim 10^{-29}$\,g/cm$^3$).
Then Eq.~(\ref{be2}) can be integrated to give
\begin{eqnarray}
\label{Phim}
e^{2\Phi (r)}=\left[ 1-\frac{2Gm(r)}{r} \right]^{-1}\,,\quad
m(r)=\int_0^r 4\pi r'^2 \rho_m\,{\rm d}r'\,.
\end{eqnarray}
Substituting Eqs.~(\ref{be2}) and (\ref{be3}) into
Eq.~(\ref{be1}) gives
\begin{eqnarray}
\label{fieldeq}
\phi''+\left[ \frac{1+e^{2\Phi}}{r}-4\pi G r e^{2\Phi}
(\rho_m-p_m) \right] \phi'= e^{2\Phi}
\left[ V_{,\phi} +Q (\rho_m-3p_m) \right]\,.
\end{eqnarray}

We assume that the energy density is constant 
inside ($\rho_m=\rho_A$) and outside ($\rho_m=\rho_B$) 
of the spherically symmetric body with a radius $r_c$.
Strictly speaking the conserved density $\rho_m^{(c)}$
in the Einstein frame is given by $\rho_m^{(c)}=e^{-Q \phi}\rho_m$\cite{KW,KW2,TT}.
However, since the condition $Q\phi \ll 1$ holds 
in most cases of interest, we do not need to 
distinguish between $\rho_m^{(c)}$ and $\rho_m$.

Inside the spherically symmetric body ($0<r<r_c$)
we have $m(r)=4\pi r^3 \rho_A/3$ and Eq.~(\ref{Phim}) gives
\begin{eqnarray}
\label{ePhi}
e^{2\Phi (r)}=\left( 1-\frac{8\pi G}{3} \rho_A r^2 \right)^{-1}\,.
\end{eqnarray}
With the neglect of the scalar-field contributions in 
Eqs.~(\ref{be2})-(\ref{be5}) it is known that the 
background gravitational field for $0<r<r_c$
corresponds to the Schwarzschild interior solution.
In this case the pressure $p_m(r)$ inside the body relative to 
the density $\rho_A$ can be analytically expressed as
\begin{eqnarray}
\label{pm}
\frac{p_m (r)}{\rho_A}=
\frac{\sqrt{1-2(r^2/r_c^2)\Phi_c}-\sqrt{1-2\Phi_c}}
{3\sqrt{1-2\Phi_c}-\sqrt{1-2(r^2/r_c^2)\Phi_c }}
\qquad (0<r<r_c)\,,
\end{eqnarray}
where $\Phi_c$ is the gravitational potential at the
surface of body:
\begin{eqnarray}
\label{Phic}
\Phi_c \equiv \frac{GM_c}{r_c}=\frac16 \rho_A r_c^2\,.
\end{eqnarray}
Here $M_c=4\pi r_c^3 \rho_A/3$ is the mass of the 
spherically symmetric body, and in the last equality 
in Eq.~(\ref{Phic}) we have used 
units such that $G=1/8\pi$.
Equation (\ref{pm}) shows that the pressure vanishes
at the surface of the body ($p_m(r_c)=0$).

In the following we shall derive analytic solutions for 
Eq.~(\ref{be1}), under the conditions $|\Phi (r) | \ll 1$
and $|\Psi (r)| \ll 1$. We neglect the terms higher than 
the linear order in $\Phi (r)$ and $\Psi (r)$.
{}From Eqs.~(\ref{ePhi})--(\ref{Phic})
it then follows that 
\begin{eqnarray}
\Phi (r) \simeq \Phi_c \frac{r^2}{r_c^2}\,,\qquad
\frac{p_m (r)}{\rho_A} \simeq \frac{\Phi_c}{2}
\left(1-\frac{r^2}{r_c^2} \right)\,,
\quad {\rm for} \quad 0<r<r_c\,.
\end{eqnarray}
At the centre of the body we have $p_m(0)/\rho_A \simeq 
\Phi_c/2$, which shows that the effect of the pressure 
becomes important in strong gravitational backgrounds.

Outside the body we assume that the density $\rho_B$
is very much smaller than $\rho_A$ with a vanishing pressure.
Then the metric outside the body can be approximated by the
Schwarzschild exterior solution:
\begin{eqnarray}
\label{Phiout}
\Phi (r) \simeq \frac{GM}{r}=\Phi_c \frac{r_c}{r}\,,
\qquad p_m(r) \simeq 0
\quad {\rm for} \quad  r>r_c\,.
\end{eqnarray}
%

\section{Matching solutions of the chameleon scalar field}
\label{chame}

In this section we solve the scalar-field equation 
(\ref{fieldeq}) in the relativistic gravitational backgrounds
discussed in Sec.~\ref{setup}.

In the nonrelativistic gravitational background where
the pressure $p_m$ as well as the gravitational potential $\Phi_c$ 
are negligible, the effective potential for the scalar field is defined 
as \cite{KW,KW2} 
\begin{eqnarray}
\label{Veff}
V_{\rm eff} (\phi)=V(\phi)+Q\rho_m \phi\,.
\end{eqnarray}
This potential has a minimum either when (i) $V_{,\phi}<0$ and $Q>0$
or (ii) $V_{,\phi}>0$ and $Q<0$.
An example of class of potentials satisfying (i) is 
provided by the inverse power-law potentials
$V(\phi)=M^{4+n}\phi^{-n}$ ($n>0$).
Since $f(R)$ gravity corresponds to the coupling $Q=-1/\sqrt{6}$,
the effective potential $V_{\rm eff}$  has a minimum for the 
case $V_{,\phi}>0$ (as in the case of the models proposed in 
Refs.~\cite{AGPT,Li,Hu,Star,Appleby,Tsuji08}).

For constant matter densities, $\rho_A$ and $\rho_B$,
inside and outside of the body, the effective potential
(\ref{Veff}) has two minima at the field 
values $\phi_A$ and $\phi_B$ characterized 
by the conditions
\begin{eqnarray}
\label{Vphi}
& & V_{,\phi} (\phi_A)+Q \rho_A=0\,,\\
& &V_{,\phi} (\phi_B)+Q\rho_B=0\,.
\end{eqnarray}
The former corresponds to the region with a high density (interior of the body)
that gives rise to a heavy mass squared 
$m_A^2 \equiv \frac{\rd^2 V_{{\rm eff}}}{\rd\phi^2}(\phi_A)$,
whereas the latter corresponds to the lower density region
(exterior of the body) with a lighter mass squared 
$m_B^2 \equiv \frac{\rd^2 V_{{\rm eff}}}{\rd\phi^2}(\phi_B)$.

The following boundary conditions are imposed at $r=0$ and 
$r \to \infty$:
\begin{eqnarray}
\label{boundary}
\frac{{\rm d}\phi}{{\rm d} r}(r=0)=0\,,\quad
\phi (r \to \infty)=\phi_B\,.
\end{eqnarray}
We need to consider the 
potential $(-V_{\rm eff})$ in order to find the ``dynamics''
of $\phi$ with respect to $r$. This means that the effective 
potential $(-V_{\rm eff})$ has a maximum at $\phi=\phi_A$.
The field $\phi$ is at rest at $r=0$ and begins to roll
down the potential when the matter-coupling term 
$Q \rho_A$ becomes important at a radius $r_1$.
If the field value at $r=0$ is close to $\phi_A$, the field stays
around $\phi_A$ in the region $0<r<r_1$.
The body has a thin-shell if $r_1$ is close to the radius 
$r_c$ of the body. 

The position of the minimum given in Eq.~(\ref{Vphi})
is shifted in the relativistic gravitational background.
In the following we shall derive the field profile by taking 
into account the corrections coming from the gravitational 
potential. 
Inside the body, Eq.~(\ref{fieldeq}) to
the the linear order in $\Phi_c$ reduces to:
\begin{eqnarray}
\label{fieldinside}
\phi''+\frac{2}{r} \left( 1-\frac{r^2}{2r_c^2} \Phi_c \right)
\phi'-(V_{,\phi}+Q \rho_A ) \left( 1+2\Phi_c \frac{r^2}{r_c^2}
\right)+\frac32 Q \rho_A \Phi_c \left( 1-\frac{r^2}{r_c^2}
\right)=0\,.
\end{eqnarray}

In the region $0 < r<r_1$ the field derivative of the effective 
potential around $\phi=\phi_A$ may be approximated by
$\rd V_{\rm eff}/\rd \phi =V_{,\phi}+Q\rho_A \simeq m_A^2 (\phi-\phi_A)$.
The solution to Eq.~(\ref{fieldinside}) can be obtained
by writing the field as $\phi=\phi_0+\delta \phi$, where $\phi_0$
is the solution in the Minkowski background  
and $\delta \phi$ is the perturbation induced by $\Phi_c$.
At the linear order in $\delta \phi$ and $\Phi_c$ we obtain
\begin{eqnarray}
\label{phi0e}
& &\phi_0''+\frac{2}{r}\phi_0'-m_A^2 (\phi_0-\phi_A)=0\,, \\
\label{delphi}
& & \delta \phi''+\frac{2}{r}\delta \phi'-m_A^2 \delta \phi=
\Phi_c \left[ \frac{2m_A^2 r^2}{r_c^2} (\phi_0-\phi_A)
+\frac{r}{r_c^2}\phi_0' -\frac32 Q \rho_A 
\left( 1-\frac{r^2}{r_c^2} \right) \right]\,.
\end{eqnarray}
The solution to Eq.~(\ref{phi0e}) that is regular at $r=0$ is 
given by $\phi_0 (r)=\phi_A+A(e^{-m_A r}-e^{m_A r})/r$, 
where $A$ is a constant. Substituting this solution into 
Eq.~(\ref{delphi}) we obtain the following solution for $\phi (r)$:
\begin{eqnarray}
\label{phiso1}
\phi(r)&=&\phi_A+\frac{A(e^{-m_A r}-e^{m_A r})}{r} \nonumber \\
&& -\frac{A \Phi_c}{m_A r_c^2} \left[ \left(\frac13 m_A^2 r^2
-\frac14 m_A r-\frac14 +\frac{1}{8m_A r} \right)e^{m_A r}+
\left(\frac13 m_A^2 r^2 +\frac14 m_A r-\frac14 -\frac{1}{8m_A r} 
\right)e^{-m_A r} \right] \nonumber \\
&& -\frac{3Q \rho_A \Phi_c}{2m_A^4 r_c^2}
\left[ m_A^2 (r^2-r_c^2)+6 \right]\,, \qquad \qquad (0<r<r_1)\,.
\end{eqnarray}
One can easily show that this solution satisfies the first of 
the boundary conditions (\ref{boundary}).

In the region $r_1<r<r_c$ the field $|\phi(r)|$ evolves 
towards larger values with increasing $r$.
Since $|V_{,\phi}| \ll |Q \rho_A|$ in this regime one has
$\rd V_{\rm eff}/\rd \phi \simeq Q \rho_A$.
In this case $\phi_0$ and $\delta \phi$ satisfy 
\begin{eqnarray}
\label{phi02}
& &\phi_0''+\frac{2}{r}\phi_0'-Q \rho_A=0\,, \\
\label{delphi2}
& & \delta \phi''+\frac{2}{r}\delta \phi'
=\Phi_c \left[ \frac{r}{r_c^2} \phi_0'
-\frac12 Q \rho_A \left( 3-7\frac{r^2}{r_c^2} \right)
\right]\,.
\end{eqnarray}
We then find the following solution 
\begin{eqnarray}
\label{phiso2}
\phi(r)=-\frac{B}{r} \left(1-\Phi_c \frac{r^2}{2r_c^2}
\right)+C+\frac16 Q \rho_A r^2 \left( 1-\frac32 \Phi_c
+\frac{23}{20} \Phi_c \frac{r^2}{r_c^2} \right)\,,\qquad
(r_1<r<r_c)\,,
\end{eqnarray}
where $B$ and $C$ are constants.

The field acquires sufficient kinetic energy in the thin-shell regime,
in order to allow it to climb up the potential hill towards 
larger absolute values in the region outside the body.
As long as the kinetic energy of the field dominates over its
potential energy, the right hand side of Eq.~(\ref{fieldeq}) 
can be neglected relative to its left hand side. 
Also the term that includes $\rho_m$ and $p_m$
in the square bracket on the left hand side of Eq.~(\ref{fieldeq})
can be neglected relative to the term $(1+e^{2\Phi})/r$.
Using Eq.~(\ref{Phiout}), the field equation reduces to 
\begin{eqnarray}
\label{phioutside}
\phi''+\frac{2}{r} \left( 1+\frac{GM}{r} \right)\phi' \simeq 0\,.
\end{eqnarray}
The solution to this equation is 
\begin{eqnarray}
\label{phiso3}
\phi (r)=\phi_B+\frac{D}{r} \left(1+\frac{GM}{r} \right) \qquad
(r>r_c)\,,
\end{eqnarray}
where $D$ is a constant. Note that here we have used the second boundary 
condition in Eq.~(\ref{boundary}).

Having obtained the solutions (\ref{phiso1}), (\ref{phiso2}) and 
(\ref{phiso3}) in the three regions inside and outside the central body,
we proceed to match these solutions at $r=r_1$ and $r=r_c$. 
The thin-shell corresponds to the region defined by
\begin{eqnarray}
\Delta r_c \equiv r_c-r_1 \ll r_c\,,
\end{eqnarray}
namely $\Delta r_c/r_c \ll 1$.

It is possible to satisfy the local gravity constraints as long as the 
field inside the body is sufficiently massive, i.e.,  
$m_A r_c \gg 1$ (or $m_A r_1 \gg 1$).
For example, in the case of the Earth with the class 
of potentials $V(\phi)=M^{4+n}\phi^{-n}$, we have the constraints 
$m_A r_c \gtrsim 10^9$, for $n=1$, 
and $m_A r_c \gtrsim 10^7$, for $n=2$, from the experimental tests 
of the equivalence principle \cite{TT}.
Since the mass $m_A$ becomes larger in higher density regions, 
the quantity $m_A r_c$ inside a strong gravitational body becomes
even larger than in the case of the Sun or the Earth.
We use the approximation that $e^{-m_A r_1}$
is negligible relative to $e^{m_A r_1}$ in Eq.~(\ref{phiso1}).

We recall that $\Phi_c={\cal O}(10^{-2})$-${\cal O}(10^{-1})$ 
for neutron stars, $\Phi_c={\cal O}(10^{-4})$-${\cal O}(10^{-2})$ 
for white dwarfs, $\Phi_c={\cal O}(10^{-6})$ for the Sun and 
$\Phi_c={\cal O}(10^{-9})$ for the Earth.
In the following we shall use linear expansions in terms
of the three parameters $\Delta r_c/r_c$, $\Phi_c$ and 
$1/(m_A r_c)$ (or $1/(m_A r_1)$).
We drop terms of higher order in these parameters 
relative to 1. We caution that our approximation loses 
its accuracy under the really strong gravitational backgrounds
with $\Phi_c \gtrsim {\cal O} (0.1)$.

Using the continuity of $\phi(r)$ and $\phi' (r)$ 
at $r=r_1$ and $r=r_c$ we obtain
\begin{eqnarray}
\label{eq1}
& & A \frac{e^{m_A r_1}}{r_1}
\left[ 1+ \frac{m_A r_1^3 \Phi_c}{3r_c^2} \left(
1-\frac{3}{4m_A r_1} \right) \right]
-B\frac{1}{r_1} \left( 1-\Phi_c \frac{r_1^2}{2r_c^2} \right)+C=\phi_A
-\frac{Q \rho_A r_1^2}{6} \left( 1-\frac32 \Phi_c+\frac{23}{20}\Phi_c
\frac{r_1^2}{r_c^2} \right)\,,\\
\label{eq2}
& & A \frac{m_A e^{m_A r_1}}{r_1}
\left[ 1+\frac{m_A r_1^3 \Phi_c}{3r_c^2} \left( 1
+\frac{5}{4m_A r_1} \right)-\frac{1}{m_A r_1} \right]
+B \frac{1}{r_1^2} \left( 1+\Phi_c \frac{r_1^2}{2r_c^2} \right) 
=-\frac16 Q \rho_A r_1 \left( 2-3\Phi_c+\frac{23}{5}\Phi_c \frac{r_1^2}{r_c^2}
\right), \nonumber \\ \\
\label{eq3}
& & B \frac{1}{r_c} \left( 1-\frac{\Phi_c}{2} \right)-C+D \frac{1}{r_c}
(1+\Phi_c)=-\phi_B+\frac16 Q \rho_A r_c^2 \left(1 -\frac{7}{20}\Phi_c \right)\,,\\
\label{eq4}
& & B\left( 1+\frac{\Phi_c}{2} \right)+D(1+2\Phi_c)=-\frac16 Q \rho_A r_c^3
\left( 2+\frac85 \Phi_c \right)\,.
\end{eqnarray}

The value of $C$ can be derived from Eqs.~(\ref{eq3}) and (\ref{eq4}) 
keeping terms to linear order in $\Phi_c$ only. Substituting 
$C$ into Eq.~(\ref{eq1}) and using Eq.~(\ref{eq2}) we can 
obtain expressions for $A$ and $B$. 
The coefficient $D$ is then obtained from Eq.~(\ref{eq4}).
Using this procedure we find
\begin{eqnarray}
\label{Asol}
&& A=\frac{1}{m_A e^{m_A r_1}}
\left[1+\Phi_c \frac{r_1^2}{4r_c^2}+
\frac{m_A r_1^3 \Phi_c}{3r_c^2} \left( 1- \Phi_c \frac{r_1^2}{2r_c^2}
\right) \right]^{-1} \Biggl[
(\phi_A-\phi_B) \left(1+\Phi_c \frac{r_1^2}{2r_c^2} \right)
+\frac12 Q \rho_A r_c^2 \left(1 -\frac{\Phi_c}{4}+\Phi_c \frac{r_1^2}{2r_c^2}
\right) \nonumber \\
&&~~~~~~-\frac12 Q \rho_A r_1^2 \left(1-\frac32 \Phi_c
+\frac{7}{4}\Phi_c \frac{r_1^2}{r_c^2} \right) \Biggr]\,,\\
\label{Bsol}
&& B=-(1-\alpha)r_1 \Biggl[ (\phi_A-\phi_B) 
\left(1+\Phi_c \frac{r_1^2}{2r_c^2} \right)+
\frac12 Q \rho_A r_c^2 
\left(1 -\frac{\Phi_c}{4}+\Phi_c \frac{r_1^2}{2r_c^2} \right)
- \frac12 Q \rho_A r_1^2 \left(1-\frac32 \Phi_c
+\frac{7}{4}\Phi_c \frac{r_1^2}{r_c^2} \right) \biggr]
 \nonumber \\
& &~~~~~~~-\frac13 Q \rho_A r_1^3 
\left(1-\frac32 \Phi_c +\frac{9}{5}\Phi_c \frac{r_1^2}{r_c^2} \right)\,,\\
\label{Csol}
&& C=\phi_B-\frac12 Q \rho_A r_c^2 \left( 1-\frac{\Phi_c}{4}
\right)\,,\\
\label{Dsol}
& & D=(1-\alpha)r_1 \Biggl[ (\phi_A-\phi_B) 
\left(1+\Phi_c \frac{r_1^2}{2r_c^2}-\frac32 \Phi_c \right)
+\frac12 Q \rho_A r_c^2 
\left(1 -\frac{7}{4}\Phi_c+\Phi_c \frac{r_1^2}{2r_c^2} \right)
- \frac12 Q \rho_A r_1^2 \left(1-3\Phi_c
+\frac{7}{4}\Phi_c \frac{r_1^2}{r_c^2} \right) \biggr]
\nonumber \\
& &~~~~~~-\frac13 Q \rho_A r_c^3 \left(1-\frac65 \Phi_c \right)
+\frac13 Q \rho_A r_1^3 \left(1-3\Phi_c+\frac95 \Phi_c \frac{r_1^2}{r_c^2} 
\right)\,,
\end{eqnarray}
where 
\begin{eqnarray}
\label{alpha}
\alpha \equiv \frac{(r_1^2/3r_c^2)\Phi_c+1/(m_A r_1)}
{1+(r_1^2/4r_c^2)\Phi_c+(m_A r_1^3 \Phi_c/3r_c^2)
(1-(r_1^2/2r_c^2)\Phi_c)}\,.
\end{eqnarray}
Since the denominator in Eq.~(\ref{alpha}) is larger than 1, 
the parameter $\alpha$ is much smaller than 1.

The distance $r_1$ is determined by the condition 
\begin{eqnarray}
\label{masssq}
m_A^2 \left[ \phi (r_1)-\phi_A \right]=Q\rho_A\,,
\end{eqnarray}
where 
\begin{eqnarray}
\label{phirap}
\phi (r_1) = \phi_A -A\frac{e^{m_A r_1}}{r_1}
\left[1+\frac{m_A r_1^3 \Phi_c}{3r_c^2}
\left(1-\frac{3}{4m_A r_1} \right)\right]
-\frac{3Q \rho_A \Phi_c}{2m_A^4 r_c^2} 
\left[ m_A^2 (r_1^2-r_c^2)+6 \right]\,.
\end{eqnarray}
Substituting Eq.~(\ref{phirap}) into Eq.~(\ref{masssq}) gives
\begin{eqnarray}
\label{Aano}
A=-\frac{Q \rho_A r_1}{m_A^2 e^{m_A r_1}}
\left( 1+\frac{m_A r_1^3 \Phi_c}{3r_c^2}
-\frac{\Phi_c}{4} \frac{r_1^2}{r_c^2} \right)^{-1}\,.
\end{eqnarray}
{}From Eqs.~(\ref{Asol}) and (\ref{Aano}) we then obtain
\begin{eqnarray}
\label{phiAB}
\phi_A-\phi_B=-Q \rho_A r_c^2
\left[ \frac{\Delta r_c}{r_c} \left(1+\Phi_c
-\frac12 \frac{\Delta r_c}{r_c} \right)+
\frac{1}{m_A r_c} \left( 1-\frac{\Delta r_c}{r_c} 
\right)(1-\beta)\right]\,,
\end{eqnarray}
where 
\begin{eqnarray}
\label{beta}
\beta \equiv \frac{(m_A r_1^3 \Phi_c/3r_c^2)(r_1^2/r_c^2)\Phi_c}
{1+(m_A r_1^3 \Phi_c/3r_c^2)
-(r_1^2/4r_c^2)\Phi_c}\,.
\end{eqnarray}
Note that $\beta \ll 1$.

The thin-shell parameter introduced in Refs.~\cite{KW,KW2}
is in this case given by 
\begin{eqnarray}
\label{thinpara}
\epsilon_{\rm th}
&\equiv& \frac{\phi_B-\phi_A}{6Q\Phi_c} \\
\label{thinpara2}
&=& \frac{\Delta r_c}{r_c} \left(1+\Phi_c
-\frac12 \frac{\Delta r_c}{r_c} \right)+
\frac{1}{m_A r_c} \left( 1-\frac{\Delta r_c}{r_c} 
\right)(1-\beta)\,.
\end{eqnarray}
To the first-order in expansion parameters one has 
$\epsilon_{\rm th}=\Delta r_c/r_c+1/(m_A r_c)$, 
which is identical to the corresponding value derived 
in the Minkowski background \cite{TT}.
The effect of the gravitational potential appears as a
second-order term to the thin-shell parameter.
Substituting Eq.~(\ref{phiAB}) into Eq.~(\ref{Dsol}),
we obtain the following approximate solution
\begin{eqnarray}
D \simeq -6Q \Phi_c r_c \left[ \frac{\Delta r_c}{r_c}
\left( 1-\frac{\Delta r_c}{r_c} \right)
+\frac{1}{m_A r_c} \left(1-
2\frac{\Delta r_c}{r_c}-\Phi_c-\alpha-\beta
 \right) \right]\,,
\end{eqnarray}
where we have carried out a linear expansion in terms 
of $\alpha$, $\beta$, $\Delta r_c/r_c$ and $\Phi_c$.
The solution outside the body is then given by 
\begin{eqnarray}
\label{phiso}
\phi(r) \simeq \phi_B-2Q_{\rm eff}\frac{GM}{r}
\left(1+\frac{GM}{r} \right)\,,
\end{eqnarray}
where the effective coupling is
\begin{eqnarray}
\label{Qeff2}
Q_{\rm eff} =
3Q \left[ \frac{\Delta r_c}{r_c}
\left( 1-\frac{\Delta r_c}{r_c} \right)
+\frac{1}{m_A r_c} \left(1-
2\frac{\Delta r_c}{r_c}-\Phi_c-\alpha-\beta
 \right) \right]\,.
\end{eqnarray}
To leading-order this gives
$Q_{\rm eff}=3Q (\Delta r_c/r_c+1/m_A r_c)=3Q \epsilon_{\rm th}$, 
which agrees with the corresponding
result in the Minkowski background \cite{TT}.
Thus provided that $\epsilon_{\rm th} \ll 1$, the effective coupling 
$Q_{\rm eff}$ becomes much smaller than the bare coupling $Q$.
The gravitational potential $\Phi_c$ appears as a next-order term.
As can be seen from Eq.~(\ref{Qeff2}) the presence of the 
gravitational potential $\Phi_c$ leads to a small decrease
in $Q_{\rm eff}$ compared to the nonrelativistic 
gravitational background.

\section{The field profile}
\label{profile}

In this section we shall discuss the analytical field profile 
derived in the previous section in more details.
The coefficient $A$ and the field difference
$\phi_A-\phi_B$ are determined by fixing the value of $r_1$,
see Eqs.~(\ref{Aano}) and (\ref{phiAB}).
{}From Eqs.~(\ref{phiso1}), (\ref{phiso2}), (\ref{phiso3}), 
(\ref{Bsol})-(\ref{Dsol}), (\ref{Aano}) and (\ref{phiAB}) 
the thin-shell field profile is given by 
\begin{eqnarray}
\label{phithin1}
\phi (r) &=& \phi_A+\frac{Q \rho_A}{m_A^2e^{m_A r_1}}
\frac{r_1}{r} \left( 1+\frac{m_A r_1^3 \Phi_c}{3r_c^2}
-\frac{\Phi_c r_1^2}{4r_c^2}  \right)^{-1}
(e^{m_A r}-e^{-m_A r})+\frac{3Q \rho_A \Phi_c}{2m_A^2} 
\left[ 1-\frac{r^2}{r_c^2}-\frac{6}{(m_A r_c)^2} \right]
\nonumber \\
& & +\frac{\Phi_c r_1}{m_A r_c^2} 
\frac{Q \rho_A}{m_A^2e^{m_A r_1}}
 \left( 1+\frac{m_A r_1^3 \Phi_c}{3r_c^2}
-\frac{\Phi_c r_1^2}{4r_c^2} \right)^{-1} \nonumber 
\\
& &\times \biggl[ \left(\frac13 m_A^2 r^2
-\frac14 m_A r-\frac14 +\frac{1}{8m_A r} \right)e^{m_A r}+
\left(\frac13 m_A^2 r^2 +\frac14 m_A r-\frac14 -\frac{1}{8m_A r} 
\right)e^{-m_A r} \biggr] \qquad (0<r<r_1), \nonumber \\
\\
\label{phithin2}
\phi (r) &=& \phi_A+\frac{Q\rho_A r_c^2}{6}
\left[ 6\epsilon_{\rm th}+6\tilde{B}\frac{r_1}{r}
\left( 1-\frac{\Phi_c r^2}{2r_c^2} \right)
-3\left(1-\frac{\Phi_c}{4}\right)+
\left( \frac{r}{r_c} \right)^2
\left( 1-\frac32 \Phi_c+\frac{23\Phi_c r^2}{20r_c^2}
\right) \right]
\quad (r_1<r<r_c),\nonumber \\
\\
\label{phithin3}
\phi (r) &=& \phi_A+Q\rho_A r_c^2 \left[ 
\epsilon_{\rm th} -\tilde{D} \frac{r_c}{r}
\left( 1+\Phi_c \frac{r_c}{r} \right) \right]
\qquad (r>r_c),
\end{eqnarray}
where 
\begin{eqnarray}
\tilde{B} \equiv -\frac{B}{Q \rho_A r_c^2 r_1}
&=& (1-\alpha)\left[ -\epsilon_{\rm th}  \left(1+
\frac{\Phi_c r_1^2}{2r_c^2} \right)
+\frac12 \left(1 -\frac{\Phi_c}{4}
+\frac{\Phi_c r_1^2}{2r_c^2} \right)
- \frac{r_1^2}{2r_c^2} \left(1-\frac32 \Phi_c
+\frac{7 \Phi_c r_1^2}{4r_c^2} \right) \right] \nonumber \\
& &+\frac{r_1^2}{3r_c^2} \left(1-\frac32 \Phi_c+\frac{9\Phi_c r_1^2}
{5r_c^2} \right)\,,\\
\tilde{D} \equiv -\frac{D}{Q \rho_A r_c^3}&=&
(1-\alpha)\left[ \epsilon_{\rm th}  \frac{r_1}{r_c}
\left(1+\frac{\Phi_c r_1^2}{2r_c^2}-\frac{3\Phi_c}{2} \right)
-\frac{r_1}{2r_c}\left(1-\frac{7}{4}\Phi_c+\frac{\Phi_c r_1^2}{2r_c^2} \right)+\frac{r_1^3}{2r_c^3} \left( 1-3\Phi_c+\frac{7\Phi_c r_1^2}
{4r_c^2} \right) \right] \nonumber \\
& &+\frac13 \left( 1-\frac65 \Phi_c \right)
-\frac{r_1^3}{3r_c^3} \left( 1-3\Phi_c+\frac{9\Phi_c r_1^2}{5r_c^2}
\right)\,.
\end{eqnarray}
Note that the field profile given in 
Eqs.~(\ref{phithin1})-(\ref{phithin3}) 
has been derived without specifying the form of the potential.
While the term $\rho_A$ in Eqs.~(\ref{phithin1})-(\ref{phithin3})
can be replaced by $6\Phi_c/r_c^2$, we have chosen not to do this 
so that the field profile in the Minkowski background can be 
simply recovered by setting $\Phi_c=0$.

In the following we shall consider in details the field profile in three
regions: (i) $0<r<r_2 \equiv 1/m_A$, (ii) $r_2<r<r_1$
and (iii) $r>r_1$, and discuss the case $Q>0$ for simplicity. 

\subsection{The region $0<r<r_2$}

Deep inside the body where the distance $r$ 
satisfies the condition $r \ll 1/m_A$,
Eq.~(\ref{phithin1}) gives the following approximate
field value and its derivative with respect to $r$:
\begin{eqnarray}
\label{phiv1}
& &\phi(r) \simeq \phi_A+\frac{2Q\rho_A r_1}
{m_A e^{m_A r_1}}
\left( 1+\frac{m_A r_1^3 \Phi_c}{3r_c^2}
-\frac{\Phi_c r_1^2}{4r_c^2}  \right)^{-1}
\left[ 1+\frac16 (m_A r)^2+\frac{\Phi_c}{2(m_A r_c)^2} \right]
+\frac{3Q\rho_A \Phi_c}{2m_A^2}
\left[1-\frac{r^2}{r_c^2}-\frac{6}{(m_A r_c)^2}
\right],\nonumber \\
\\
\label{phiv2}
& &\phi'(r) \simeq Q \rho_A r_c^2
\left[ \frac{2m_A r_1}{3e^{m_A r_1}}
\left( 1+\frac{m_A r_1^3 \Phi_c}{3r_c^2}
-\frac{\Phi_c r_1^2}{4r_c^2}  \right)^{-1}
-\frac{3\Phi_c}{(m_A r_c)^2} \right]
\frac{r}{r_c^2}\,.
\end{eqnarray}

In the Minkowski background ($\Phi_c=0$) we have 
$\phi(0) \simeq \phi_A+2Q\rho_A r_1/(m_A e^{m_A r_1})$
and $\phi'(r)>0$ (where $r \neq 0$).
Hence the field rolls down the potential toward larger $\phi$
with increasing $r$.
In the presence of the gravitational potential $\Phi_c$, 
the derivative $\phi'(r)$ can be negative 
depending on model parameters.
Using the approximation $r_c \simeq r_1$ the condition that 
$\phi'(r)<0$ translates into
\begin{eqnarray}
\label{phinecon}
\frac{(m_A r_1)^3}{e^{m_A r_1}}-\frac32\Phi_c^2 m_A r_1<
\frac92 \Phi_c \left(1-\frac{\Phi_c}{4} \right)\,.
\end{eqnarray}
When $\Phi_c>0.253$ this is automatically 
satisfied for all (positive) $m_A r_1$.
When $\Phi_c=10^{-1}, 10^{-6}, 10^{-9}$, the 
condition (\ref{phinecon}) is satisfied for 
$m_A r_1>6$, $m_A r_1>22$ and $m_A r_1>29$, 
respectively. Hence in most realistic cases
where $m_A r_1 \gg 1$ we have $\phi'(r)<0$ in the region $0<r<r_2$.
Interestingly this property persists even in the weak 
gravitational backgrounds, such as those of 
the Sun or the Earth.

The evolution towards the smaller $\phi$ region is due
to the effects of the relativistic pressure $p_m$ since 
the last term on the right hand side of Eq.~(\ref{phiv1})
originates from the pressure.
Compared to the case of the Minkowski spacetime, 
the presence of the pressure term leads to 
the shift of the field $\phi(0)$ 
towards a larger value. When this effect of the pressure 
dominates over the rolling down effect 
along the potential, we have 
\begin{eqnarray}
\label{phifirst}
\phi (r) \simeq \phi_A+\frac{3Q\rho_A \Phi_c}{2m_A^2}
\left[1-\frac{r^2}{r_c^2}-\frac{6}{(m_Ar_c)^2}
\right]\,.
\end{eqnarray}
This shows that the field decreases from 
$\phi(0) \simeq \phi_A+\frac{3Q\rho_A \Phi_c}{2m_A^2}
\left[1-6/(m_Ar_c)^2
\right]$ to 
$\phi(r_2) \simeq \phi_A+\frac{3Q\rho_A \Phi_c}{2m_A^2}
\left[1-7/(m_Ar_c)^2
\right]$ with increasing $r$.

\subsection{The region $r_2<r<r_1$}

In the region $r_2<r<r_1$ we have the following approximate 
solutions from Eq.~(\ref{phithin1}):
\begin{eqnarray}
\label{phiw1}
\phi(r) &\simeq& \phi_A+
\frac{Q \rho_A }{m_A^2} e^{m_A (r-r_1)}\,
\frac{r_1/r+m_A \Phi_c r_1 r^2/3r_c^2-\Phi_c r_1 r/4r_c^2}
{1+m_A r_1^3\Phi_c/3r_c^2-\Phi_c r_1^2/4r_c^2}+
\frac{3Q\rho_A \Phi_c}{2m_A^2}
\left[1-\frac{r^2}{r_c^2}-\frac{6}{(m_A r_c)^2}
\right]\,,\\
\label{phiw2}
\phi'(r) &\simeq& 
\frac{Q\rho_A}{m_A^2} \left( 1+\frac{m_A r_1^3 \Phi_c}{3r_c^2}
-\frac{\Phi_c r_1^2}{4r_c^2} \right)^{-1} m_A e^{m_A (r-r_1)} \nonumber \\
& &\times
\left[ \frac{r_1}{r}+\frac{\Phi_c m_A r_1 r^2}{3r_c^2}
-\frac{\Phi_c r_1r}{4r_c^2}-\frac{1}{m_A r}
\left(\frac{r_1}{r}-\frac{2\Phi_c m_A r_1 r^2}{3r_c^2}
+\frac{\Phi_c r_1 r}{4r_c^2} \right) \right]
-\frac{3Q\rho_A \Phi_c}{(m_A r_c)^2}r\,.
\end{eqnarray}
In Eq.~(\ref{phiw1}) we have taken into account the terms 
$(m_A^2r^2/3-m_Ar/4)e^{m_A r}$
in the last square bracket of Eq.~(\ref{phithin1}).
If $\phi'(r)<0$ at $r=r_2$, the field derivative needs to 
change its sign from negative to positive 
at the distance $r_3$ (i.e., $\phi'(r_3)=0$).

{}From Eq.~(\ref{phiw2}) the condition $\phi'(r_3)=0$
translates to
\begin{eqnarray}
\label{r3es}
& &m_A r_c e^{m_A (r_3-r_1)}
\left[ \frac{r_1}{r_3}+\frac{\Phi_c m_A r_1 r_3^2}{3r_c^2}
-\frac{\Phi_c r_1r_3}{4r_c^2}-\frac{1}{m_A r_3}
\left(\frac{r_1}{r_3}-\frac{2\Phi_c m_A r_1 r_3^2}{3r_c^2}
+\frac{\Phi_c r_1 r_3}{4r_c^2} \right) \right] \nonumber \\
& &=3\Phi_c \frac{r_3}{r_c}
\left( 1+\frac{m_A r_1^3 \Phi_c}{3r_c^2}
-\frac{\Phi_c r_1^2}{4r_c^2} \right).
\end{eqnarray}
In Table I we show the values of $r_3/r_1$ 
under the approximation $r_c \simeq r_1$
for several different choices of $\Phi_c$ and $m_A r_1$.
Clearly $r_3/r_1$ gets larger with increasing $\Phi_c$ 
and $m_A r_1$. 
When $\Phi_c=10^{-6}$ and $\Phi_c=10^{-9}$
with $m_A r_1=10$ the field satisfies the condition 
$\phi'(r)>0$ in the region $r_2<r<r_1$.
Meanwhile, when $\Phi_c=10^{-1}$ and $m_A r_1=10$,
the sign change of $\phi'(r)$ occurs at $r_3/r_1=0.55$.

\begin{table*}[t]
\begin{center}
\begin{tabular}{|c|c|c|c|}
\hline
~ & $m_A r_1=10$ & $m_A r_1=10^2$  & $m_A r_1=10^3$ \\
\hline
$\Phi_c=10^{-9}$ & $\phi'(r)>0$ & 0.75 & 0.97 \\
\hline
$\Phi_c=10^{-6}$ & $\phi'(r)>0$ & 0.82 & 0.98  \\
\hline
$\Phi_c=10^{-1}$ & 0.55& 0.94 & 0.99  \\
\hline
\end{tabular}
\end{center}
\caption[staghost]{\label{table1} 
The values of $r_3/r_1$ at which $\phi'(r_3)=0$ under the 
approximation $r_c \simeq r_1$.
It is clear that $r_3/r_1$ increases for 
larger $\Phi_c$ and $m_A r_1$.
In the cases $\Phi_c=10^{-9}$ and $\Phi_c=10^{-6}$ with 
$m_A r_1=10$ the field derivatives $\phi'(r)$ are positive 
in the region $1/m_A<r<r_1$.
}
\end{table*}

In the region $r \gtrsim r_3$ the second term on the right hand side 
of Eq.~(\ref{phiw1}) dominates over the third one, 
giving the following solution
\begin{eqnarray}
\phi(r) \simeq \phi_A+
\frac{Q \rho_A }{m_A^2} e^{m_A (r-r_1)}\,
\frac{r_1/r+m_A \Phi_c r_1 r^2/3r_c^2-\Phi_c r_1 r/4r_c^2}
{1+m_A r_1^3\Phi_c/3r_c^2-\Phi_c r_1^2/4r_c^2}\,.
\end{eqnarray}
At $r=r_1$ we have $\phi(r_1)=\phi_A+Q\rho_A/m_A^2$,
as required by Eq.~(\ref{masssq}).
Note that $|\phi(r_1)-\phi_A|$ is larger than $|\phi(0)-\phi_A|$
by a factor of $2/(3\Phi_c)$.

\subsection{The region $r>r_1$}

In the region $r_1<r<r_c$ the field $\phi$ grows because of the 
dominance of the last term in Eq.~(\ref{phithin2}).
The field value at the surface of the body can be estimated as
\begin{eqnarray}
\phi(r_c) \simeq \phi_A+Q\rho_A r_c^2
\left[\frac{1}{m_A r_c}\alpha+\frac12 \left( \frac{\Delta r_c}{r_c}
\right)^2+\frac{1}{m_A r_c} \frac{\Delta r_c}{r_c} \right] 
\simeq \phi_A+\frac{Q\rho_A}{m_A^2}
\left[ 1+\frac12 \left( m_A r_c \frac{\Delta r_c}{r_c} \right)^2
+m_A r_c \frac{\Delta r_c}{r_c} \right]\,,
\end{eqnarray}
where in the second approximate equality we have used 
the fact that $\alpha$ is of the order of $1/(m_Ar_1)$.
Obviously $\phi (r_c)$ is larger than $\phi (r_1)$.
If the condition 
\begin{eqnarray}
\label{delrcon}
\frac{\Delta r_c}{r_c} \gg \frac{1}{m_A r_c}\,,
\end{eqnarray}
is satisfied, we have that $\phi (r_c) \gg \phi (r_1)$.
In this case the field acquires a sufficient amount of 
kinetic energy so that the following condition 
is satisfied at $r=r_c$:
\begin{eqnarray}
\label{phiapeq}
\phi''+\frac{2}{r}\phi'  
\simeq Q\rho_A \gg |V_{,\phi}|\,.
\end{eqnarray}
We recall that outside the body the density 
$\rho_A$ sharply drops down to $\rho_B$.
Hence only the potential-dependent term remains
on the right hand side of Eq.~(\ref{fieldeq}).
Under the condition (\ref{phiapeq}) 
the kinetic energy dominates over $|V_{,\phi}|$
for $r>r_c$ so that the field equation is approximately 
given by Eq.~(\ref{phioutside}).
In this case the analytic field profile 
should be trustable.

If the condition (\ref{delrcon}) is not satisfied, the field
$\phi (r_c)$ is not much different from $\phi(r_1)$.
In this case the kinetic energy of the field is not sufficiently 
large so that the term $|V_{,\phi}|$ is not 
negligible relative to $Q\rho_A$ at $r=r_c$.
In the region $r>r_c$, this can lead to 
the pullback of the field because
the kinetic energy is not large enough for the field to climb
up the potential hill.
In fact we have numerically confirmed this behaviour
in cases where $\Delta r_c/r_c$ is smaller than the order 
of $1/(m_A r_c)$.
Thus the condition (\ref{delrcon}) is important in order to 
obtain the field solution (\ref{phiso3}) outside the body.

\section{Numerical simulations}
\label{numerics}

In this section we shall numerically confirm the analytic field
profile presented in the previous section and   
discuss the validity of the approximations used to derive it.
In these numerical simulations we employ the class
of inverse power-law potentials
\begin{eqnarray}
\label{inversepo}
V(\phi)=M^{4+n}\phi^{-n} \quad (n>0)\,.
\end{eqnarray}
Although we specify the field potential to be of the form (\ref{inversepo}), 
the thin-shell field profile we will derive numercially in this section 
also holds for other potentials, such as $V(\phi)=M^4 \exp(M^n/\phi^n)$,
as can be expected from the general form 
of the Eqs.~(\ref{phithin1})-(\ref{phithin3}).
The effective potential $V_{\rm eff}$ has extrema
inside and outside the body for $Q>0$.
The field value $\phi_A$ and the mass squared 
$m_A^2$ inside the body are given by 
\begin{eqnarray}
\label{phis}
\phi_A=\left[ \frac{n}{Q} \frac{M^{4+n}}{\rho_A}
\right]^{1/(n+1)}\,, \quad
m_A^2=n(n+1) \left( \rho_A \frac{Q}{n} \right)^{(n+2)/(n+1)}
M^{-(n+4)/(n+1)}\,,
\end{eqnarray}
which lead to the following relation
\begin{eqnarray}
\phi_A=\frac{(n+1)Q\rho_A r_c^2}{(m_A r_c)^2}\,.
\end{eqnarray}
The field value $\phi_B$ and the mass $m_B$ can be obtained
by replacing $\rho_A$ for $\rho_B$ in Eq.~(\ref{phis}).

We introduce a dimensionless field $\varphi$ defined by 
\begin{eqnarray}
\varphi \equiv \phi/\phi_A\,.
\end{eqnarray}
From Eqs.~(\ref{phithin1})-(\ref{phithin3})
the analytic thin-shell field profile for the potential (\ref{inversepo}) is given by 
\begin{eqnarray}
\label{phiini}
\varphi (r) &=& 1+\frac{1}{(n+1)e^{m_A r_1}}
\frac{r_1}{r} \left( 1+\frac{m_A r_1^3 \Phi_c}{3r_c^2}
-\frac{\Phi_c}{4} \frac{r_1^2}{r_c^2} \right)^{-1}
(e^{m_A r}-e^{-m_A r})+\frac{3\Phi_c}{2(n+1)} 
\left[ 1-\frac{r^2}{r_c^2}-\frac{6}{(m_A r_c)^2} \right]
\nonumber \\
& & +\frac{\Phi_c}{(n+1)e^{m_A r_1}}\frac{r_1}{m_A r_c^2}
 \left( 1+\frac{m_A r_1^3 \Phi_c}{3r_c^2}
-\frac{\Phi_c}{4} \frac{r_1^2}{r_c^2} \right)^{-1} \nonumber 
\\
& &\times \biggl[ \left(\frac13 m_A^2 r^2
-\frac14 m_A r-\frac14 +\frac{1}{8m_A r} \right)e^{m_A r}+
\left(\frac13 m_A^2 r^2 +\frac14 m_A r-\frac14 -\frac{1}{8m_A r} 
\right)e^{-m_A r} \biggr] \qquad (0<r<r_1), \nonumber \\
\\
\label{phiini2}
\varphi (r) &=& 1+\frac{(m_A r_c)^2}{n+1} \left[
\epsilon_{\rm th}+
\tilde{B} \frac{r_1}{r}
\left( 1-\frac{\Phi_c}{2} \frac{r^2}{r_c^2} \right)
-\frac12 \left( 1-\frac14 \Phi_c \right) 
+\frac16 \left( \frac{r}{r_c} \right)^2
\left(1-\frac32 \Phi_c +\frac{23 \Phi_c r^2}{20 r_c^2} \right) \right]
\quad (r_1<r<r_c), \nonumber \\
\\
\label{phiini3}
\varphi (r) &=& 1+\frac{(m_A r_c)^2}{n+1} \left[ 
\epsilon_{\rm th} -\tilde{D} \frac{r_c}{r}
\left( 1+\Phi_c \frac{r_c}{r} \right) \right]
\qquad (r>r_c)\,.
\end{eqnarray}

Introducing a dimensionless distance normalized by $r_c$:
\begin{eqnarray}
x \equiv r/r_c\,,
\end{eqnarray}
the field equations to be solved numerically take the forms
\begin{eqnarray}
\label{dx1}
& &\frac{{\rm d}^2 \varphi}{{\rm d}x^2}+\frac{2-5\Phi_c x^2+3\Phi_c x^2\,p_m/\rho_m}
{x(1-2\Phi_c x^2)} \frac{{\rm d}\varphi}{{\rm d}x}=
\frac{(m_A r_c)^2}{n+1} \frac{1}{1-2\Phi_c x^2}
\left[ 1-\frac{3p_m(r)}{\rho_A}-\frac{1}{\varphi^{n+1}} \right] \qquad (0 <x<1)\,,\\
\label{dx2}
& & \frac{{\rm d}^2 \varphi}{{\rm d}x^2}+\frac{2(1-\Phi_c/x)-3\Phi_c x^2 \rho_B/\rho_A}
{x-2\Phi_c} \frac{{\rm d}\varphi}{{\rm d}x}=
\frac{(m_A r_c)^2}{n+1} \frac{1}{1-2\Phi_c/x}
\left( \frac{\rho_B}{\rho_A}-\frac{1}{\varphi^{n+1}} \right)
\qquad (x>1)\,,
\end{eqnarray}
where $p_m(r)/\rho_A$ is given in Eq.~(\ref{pm}).
Given the occurrence of $x$ in the denominator of 
the second term of Eq.~(\ref{dx1}), the numerical solutions cannot
start from the centre of the body ($x=0$). 
Instead we start the integrations from a radius $ r=r_i$,
slightly away from the centre satisfying the condition $r_i \ll 1/m_A$.
In so doing we use the analytic solution (\ref{phiini}) with 
the field derivative 
\begin{eqnarray}
\frac{{\rm d}\varphi}{{\rm d}x} &=&\frac{1}{n+1}
\Biggl[ \frac{1}{e^{m_A r_1}} 
\left( 1+\frac{m_A r_1^3 \Phi_c}{3r_c^2}
-\frac{\Phi_c}{4} \frac{r_1^2}{r_c^2} \right)^{-1}
\frac{r_c r_1}{r^2} \biggl\{ m_A r (e^{-m_A r}+e^{m_A r})
+e^{-m_A r}-e^{m_A r} \nonumber \\
& &+\frac{r^2}{r_c^2} \Phi_c
\biggl( \left( \frac13 m_A^2 r^2+\frac{5}{12}m_Ar-\frac12
+\frac{1}{8m_A r}-\frac{1}{8m_A^2 r^2} \right)e^{m_A r}
\nonumber \\
& &-\left( \frac13 m_A^2 r^2-\frac{5}{12}m_Ar-\frac12
-\frac{1}{8m_A r}-\frac{1}{8m_A^2 r^2} \right)e^{-m_A r}
\biggr) \biggr\}-3\Phi_c \frac{r}{r_c} \Biggr]\,.
\end{eqnarray}

{}From Eq.~(\ref{phis}) we have that 
$\phi_B/\phi_A=(\rho_A/\rho_B)^{1/(n+1)}$ and hence
$\epsilon_{\rm th}=(n+1)/(m_A r_c)^2
[ \left( \rho_A/\rho_B \right)^{1/(n+1)}-1]$.
Using Eq.~(\ref{thinpara2}) we obtain the following relation
for the ratio $\rho_A/\rho_B$
\begin{eqnarray}
\label{rhoratio}
\frac{\rho_A}{\rho_B}=\left[ 1+\frac{(m_A r_c)^2}{n+1}
\left\{ \frac{\Delta r_c}{r_c}\left(1+\Phi_c
-\frac12 \frac{\Delta r_c}{r_c} \right)+
\frac{1}{m_A r_c} \left( 1-\frac{\Delta r_c}{r_c} 
\right)(1-\beta) \right\} \right]^{n+1}\,.
\end{eqnarray}
Thus specifying the values of $n$, $m_Ar_c$, $\Delta r_c/r_c$
and $\Phi_c$, allows the ratio $\rho_A/\rho_B$ to be determined.
Alternatively, given the ratio $\rho_A/\rho_B$ together with
$n$ and $\Phi_c$, allows the relationship 
between $m_A r_c$ and $\Delta r_c/r_c$ to be derived.
We note that the condition $\Delta r_c/r_c \gg 1/(m_A r_c)$
needs to be satisfied for the field
to have a sufficient kinetic energy
outside the body.

\subsection{Minkowski background ($\Phi_c=0$)}
\label{min}

Let us first consider the Minkowski background ($\Phi_c=0$).
In this case the analytic field profile is given by 
\begin{eqnarray}
\label{minso1}
\varphi (x) &=& 1+\frac{1}{(n+1)e^{m_A r_1}}
\frac{r_1}{r_c}\frac{1}{x}
(e^{m_A r_c x}-e^{-m_A r_c x})
 \qquad (0<r<r_1), \\
 \label{minso2}
\varphi (x) &=& 1+\frac{(m_A r_c)^2}{n+1}
\left[ \epsilon_{\rm th}+\frac16 (x^2-3)
+\frac{r_1^3}{3r_c^3}\frac{1}{x}-\left( 1-\frac{1}{m_A r_1}
\right) \frac{1}{m_A r_c} \frac{r_1^2}{r_c^2} \frac{1}{x}
\right]\qquad (r_1<r<r_c), \\
\label{minso3}
\varphi (x) &=& 1+\frac{(m_A r_c)^2}{n+1}
\left[ \epsilon_{\rm th}-\frac{r_1}{r_c}\frac{1}{x}
\left\{ \epsilon_{\rm th}+\frac{r_c}{6r_1}
\left(2+\frac{r_1}{r_c}\right)
\left(1-\frac{r_1}{r_c}\right)^2-\frac{1}{(m_A r_c)^2}
\right\} \right]\qquad (r>r_c)\,.
\end{eqnarray}
In deriving Eq.~(\ref{minso3}) 
we have used the relation
$\epsilon_{\rm th}+[(r_1/r_c)^2-1]/2=r_1/(m_A r_c^2)$
coming from Eq.~(\ref{thinpara}).

\begin{figure}
\begin{centering}
\includegraphics[height=3.0in,width=3.2in]{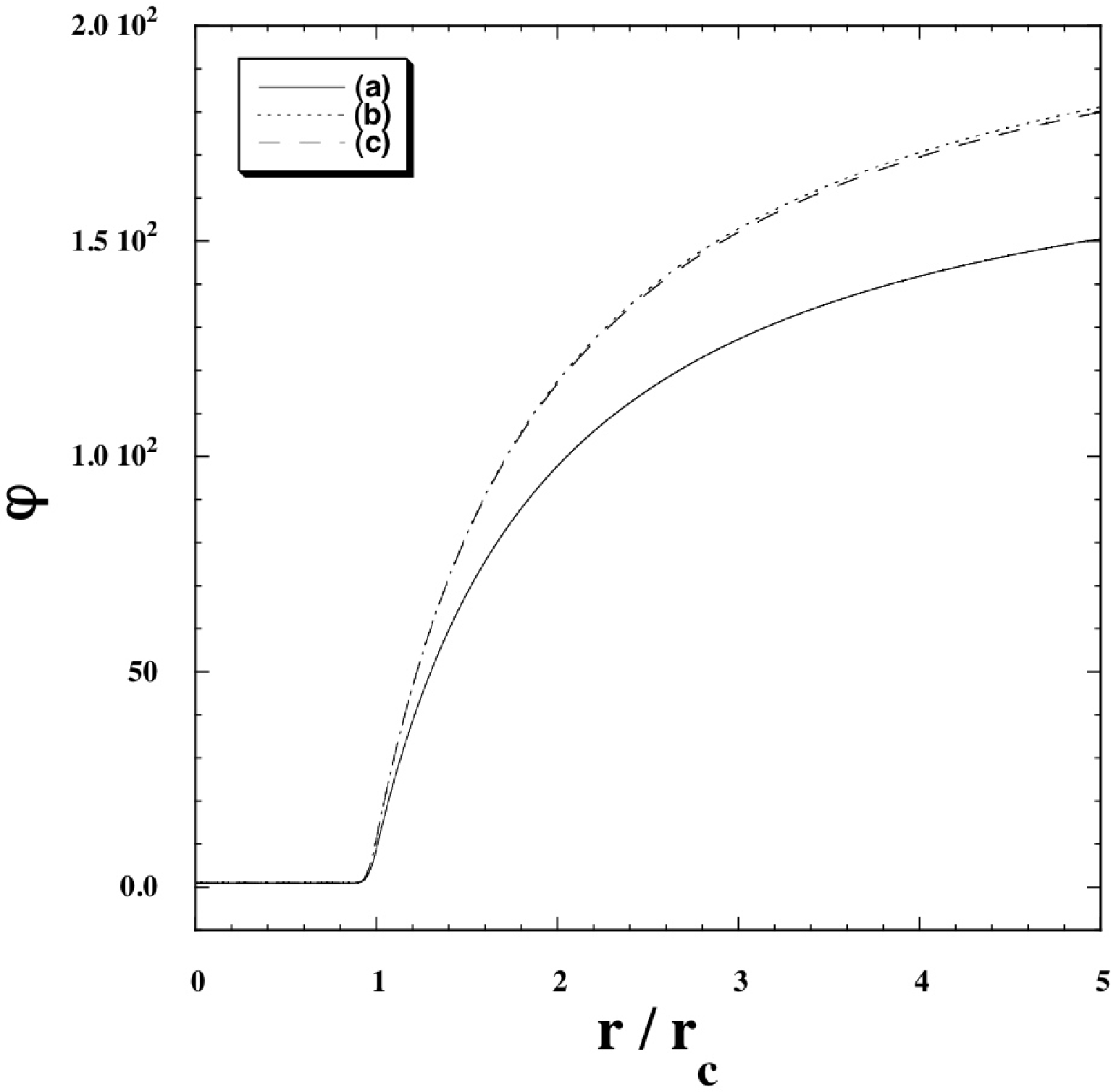}
\includegraphics[height=3.0in,width=3.2in]{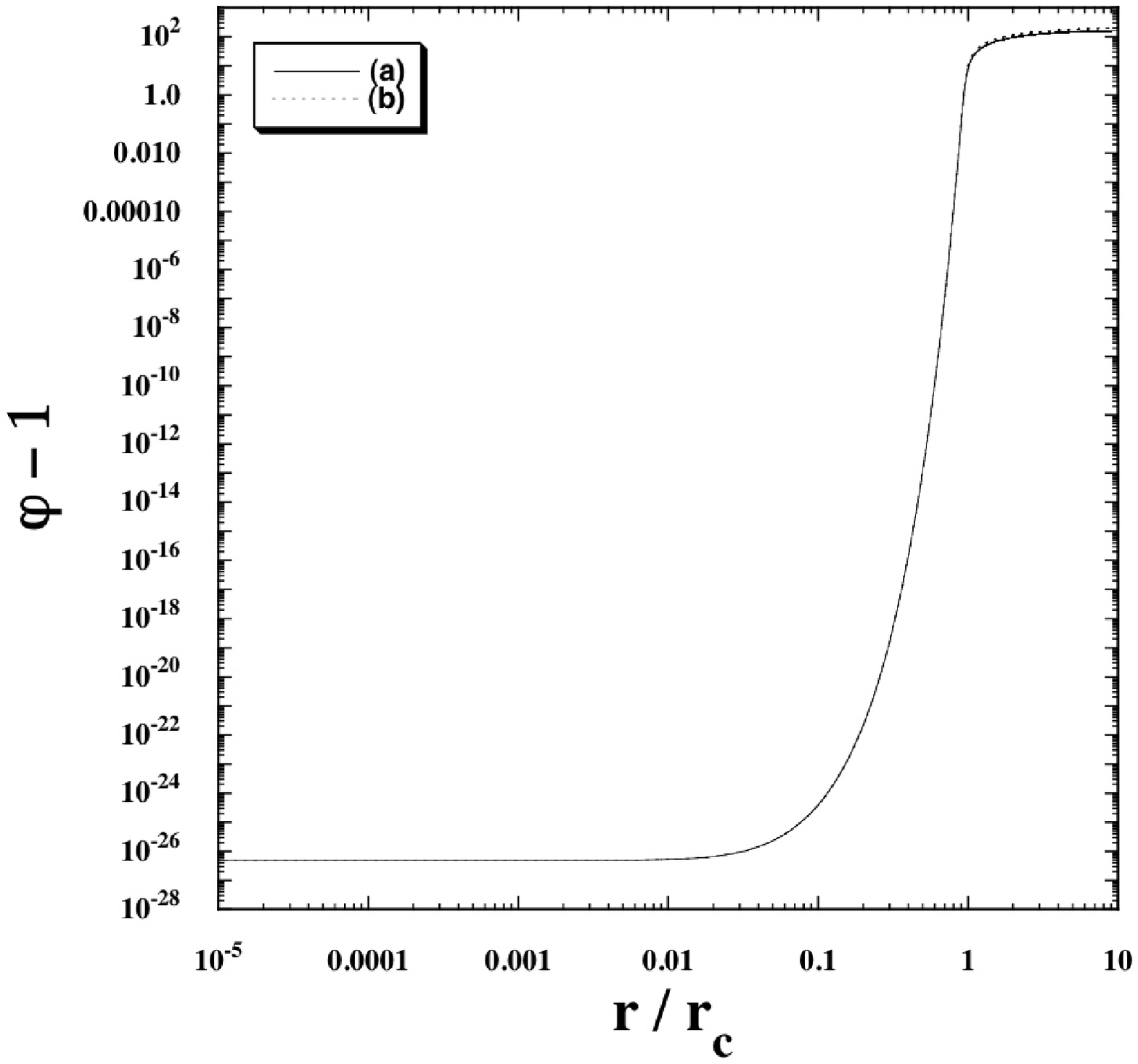}  
\par\end{centering}
\caption{The thin-shell field profile in the Minkowski background 
for $n=1$, $Q=1$, $\rho_A/\rho_B=5.0 \times 10^4$, 
and $\Delta r_c/r_c=0.08$.
This case corresponds to $1/(m_A r_c)=0.0142$ and 
$\epsilon_{\rm th}=0.0899$.
The boundary conditions for $\varphi$ and $\varphi'$ 
at $x_i=10^{-5}$ are chosen by using the analytic solution (\ref{minso1}).
In the left panel the black curve (a) shows the numerically integrated solution.
The dotted curve (b) corresponds to the analytic field profile 
given in Eqs.~(\ref{minso1})-(\ref{minso3}).
The dashed curve (c) corresponds to the numerical solution that is 
derived by solving the field equations using the approximations
$V_{{\rm eff},\phi}=m_A^2 (\phi-\phi_A)$ for $0<r<r_1$ and 
$V_{{\rm eff},\phi}=Q\rho_A$ for $r_1<r<r_c$.
While the curve (c) agrees with the curve (b) with high accuracy, 
the curve (a) deviates from the curve (b) in the region $r>r_1=0.92r_c$.
This shows that the analytic estimation that connects two solutions
at $r=r_1$ overestimates the field value outside the body
(about 20 \% larger at the distance $r=5r_c$ in this case).
The right panel is the magnified log plot of $(\varphi-1)$ in the region 
$0<r/r_c<1.4$. While the numerical solution (a) agrees well with 
the analytic solution in the region $r<r_1$, the deviation begins to 
appear in the region $r>r_1$ (in the log plot 
the deviation appears to be small).}
\label{fig1}
\end{figure}

For the parameter values $n=1$, $\rho_A=1$~g/cm$^3$, $\rho_B=10^{-4}$~g/cm$^3$ and
$\Delta r_c/r_c=0.0625$ used in the numerical simulation of Ref.~\cite{KW},
we obtain $1/(m_A r_c)=0.0200$ and 
$\epsilon_{\rm th}=0.0793$ from 
Eqs.~(\ref{rhoratio}) and (\ref{thinpara2}).
In this case the condition $\Delta r_c/r_c >1/(m_A r_c)$ is satisfied
so that the field acquires sufficient kinetic energy in the region $r_1<r<r_c$.
One can also consider the case in which the difference of  $\Delta r_c/r_c$
and $1/(m_A r_c)$ is larger.
For example, with $n=1$, $\rho_A=1$~g/cm$^3$, 
$\rho_B=2.0 \times 10^{-5}$~g/cm$^3$,
$\Delta r_c/r_c=0.08$, one has 
$1/(m_A r_c)=0.0142$ and $\epsilon_{\rm th}=0.0899$.
In Fig.~\ref{fig1} we plot the thin-shell field profile 
for this latter case by choosing the boundary conditions for $\varphi$
and $\varphi' \equiv {\rm d}\varphi/{\rm d}x$ at $x_i \equiv r_i /r_c=10^{-5}$,
using the analytic solution (\ref{minso1}).
The numerical solution (a) derived by solving Eqs.~(\ref{dx1})
and (\ref{dx2}) shows fairly good agreement with the analytic solution (b)
in the region $r<r_1=0.92r_c$. 
On the other hand the agreement is not very good in the region $r>r_1$, 
with a 20 \% difference at the distance $r=5r_c$.

\begin{figure}
\begin{centering}
\includegraphics[height=3.0in,width=3.2in]{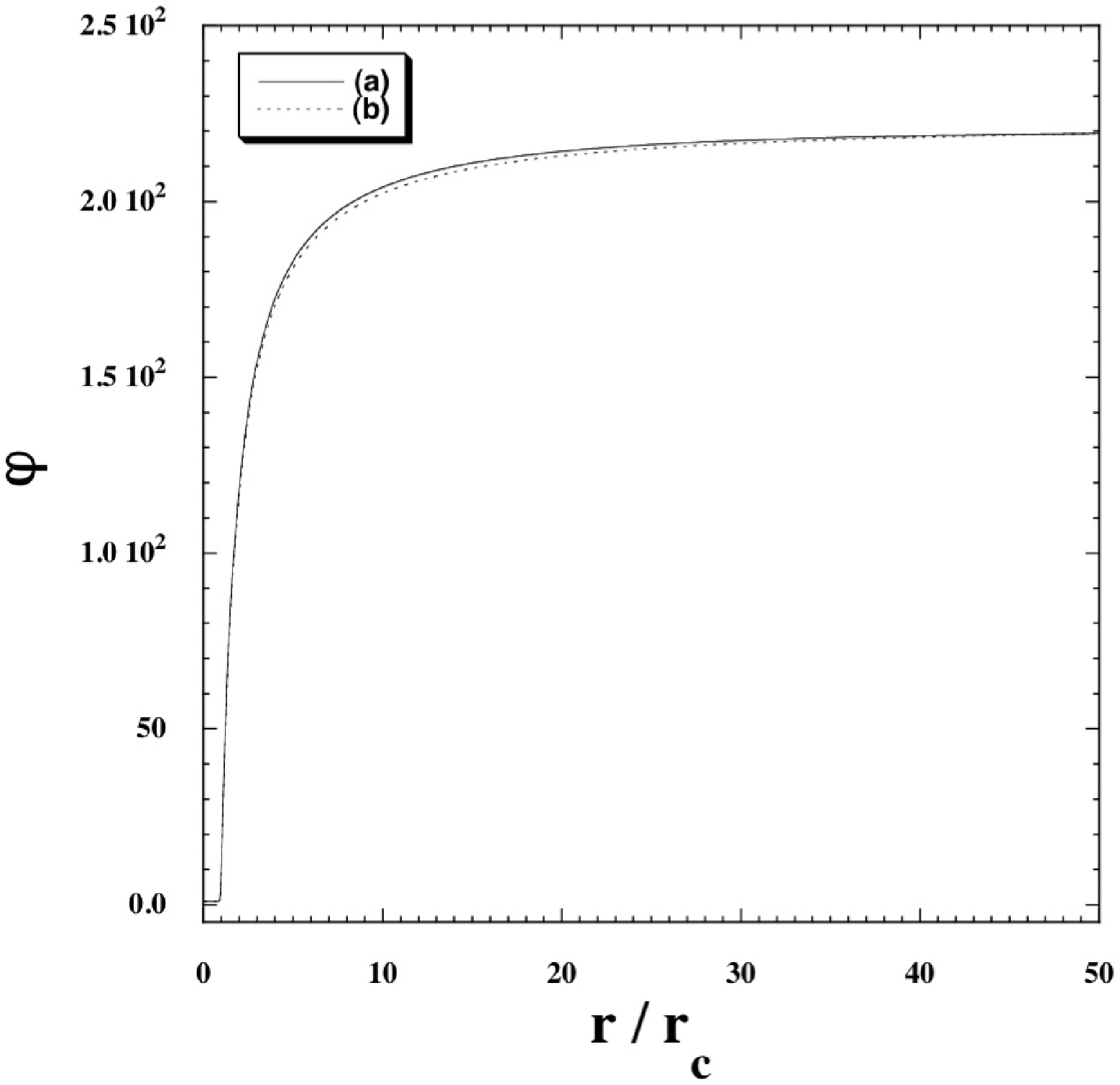}
\includegraphics[height=3.0in,width=3.2in]{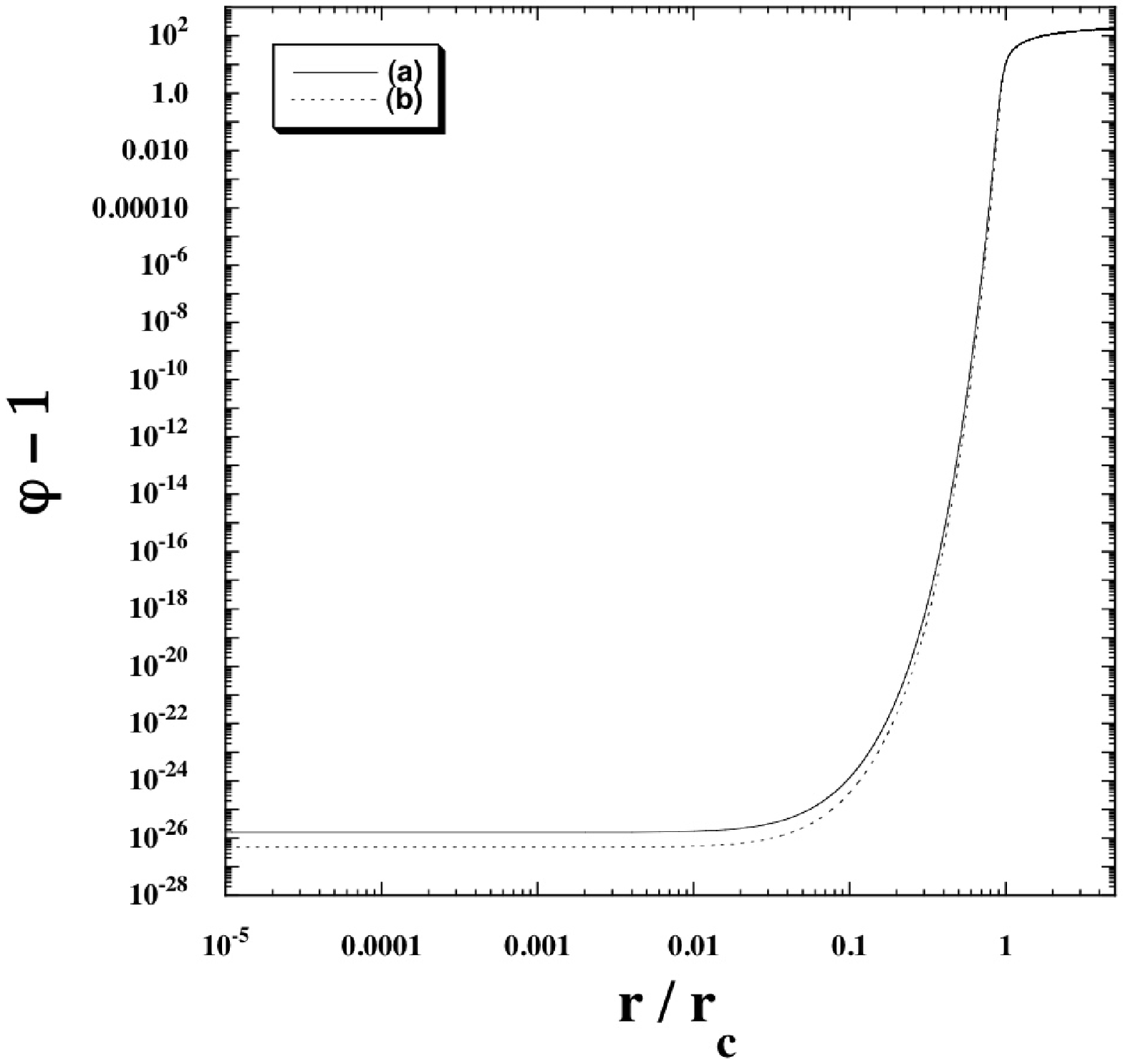}  
\par\end{centering}
\caption{The thin-shell field profile in the Minkowski background 
for the same model parameters as given in Fig.~\ref{fig1},
but with a different boundary condition 
for $\varphi$ at $x_i=10^{-5}$:
$\varphi(x_i)-1=1.630 \times 10^{-26}$ (which is larger than 
the one used in Fig.~\ref{fig1}: $\varphi(x_i)-1=4.889 \times 10^{-27}$).
The derivative $\varphi'(x)$ at $x=x_i$ is the same as
in the case of Fig.~\ref{fig1}: 
$\varphi'(x_i)=8.074 \times 10^{-29}$.
The black curve (a) and the dotted curve (b) show
the numerically integrated solution and 
the analytic field profile, respectively.
The left panel is the plot in the region $0<r/r_c<50$, whereas
the right panel is the magnified log plot of $(\varphi-1)$ in the region 
$0<r/r_c<5$. In this case the numerical solution approaches the
asymptotic field value $\varphi_B=\phi_B/\phi_A=223.6$
in the limit $r/r_c \to \infty$. In the region outside the body
the analytic solution agrees well with the numerical solution.}
\label{fig2}
\end{figure}

Inside the body the following relation holds
\begin{eqnarray}
\varphi (r_1) \simeq 1+\frac{1}{n+1}\,,\qquad
\frac{|V_{,\phi}|}{Q\rho_A} \simeq \frac{1}{\varphi^{n+1}}\,,
\end{eqnarray}
which gives $|V_{,\phi}(r_1)|/Q\rho_A \simeq 1/2$.
This shows that our analytic estimation does not hold well 
in the region around $r=r_1$.
In particular the neglect of the term $V_{,\phi}$ relative to 
$Q\rho_A$ in the region $r_1<r<r_c$ gives rise
to an error compared to the numerical simulation
including this term.
We find that the field value numerically obtained in the region 
$r_1<r<r_c$ is smaller than the analytic value given in 
Eq.~(\ref{minso2}).
This leads to the smaller field derivative $\varphi'(x)$
at the surface of the body ($x=1$).
In the numerical solution presented in Fig.~\ref{fig1}
the numerical value of $\varphi'(x=1)$ is different from its
corresponding analytic value by about 18 \%.
This difference is inherited by the field profile
outside the body.
 
In  Fig.~\ref{fig1} we also plot the numerical solution (c) 
derived by solving the field equation
with the approximation $V_{{\rm eff},\phi}=m_A^2 (\phi-\phi_A)$ 
for $0<r<r_1$ and $V_{{\rm eff},\phi}=Q\rho_A$ for $r_1<r<r_c$.
We find that the solution (c) agrees well with the analytic solution (b).
This shows that the reason for the discrepancy between the solutions
(a) and (b) is due to the fact that the matching 
of two analytic solutions at $r=r_1$ overestimates the field values
and their derivatives in the region $r_1<r<r_c$.
We have also tried other model parameters and 
have found that this property holds generally.

If we take boundary conditions with larger values of $\varphi(x)$
or $\varphi'(x)$ than those estimated by Eq.~(\ref{minso1})
around the centre of the body,
it is possible to obtain a field profile outside the body
that is close to the analytic estimation (\ref{minso3}).
In Fig.~\ref{fig2} we show the numerical solution corresponding
to the same model parameters as given in Fig.~\ref{fig1} but with a boundary 
condition for the field that is larger than the one given by the analytic solution
(\ref{minso1}). As can be seen in
this case the numerical solution outside 
the body agrees well with the analytic solution 
(\ref{minso3}). Note that the field approaches the asymptotic value 
$\varphi_B \equiv \phi_B/\phi_A=223.6$, estimated analytically using the relation
$\varphi_B=(\rho_A/\rho_B)^{1/(n+1)}$.

The above results show that the analytic solution is useful 
to find boundary conditions in order to determine the thin-shell field profile.
If we choose the field value to be slightly larger than the one 
estimated by Eq.~(\ref{minso1}) around the centre of the body, 
we are able to find a numerical solution outside the body that 
is close to the analytic solution (\ref{minso3}).

\subsection{The relativistic gravitational background ($\Phi_c \neq 0$)}
\label{relativistic}

We shall proceed to the case of the relativistic gravitational background.
As we already explained in Sec.~\ref{profile}, the presence of 
a relativistic pressure is important around the centre of the body.
This relativistic pressure gives rise to a force 
against the driving force that comes from the slope of the field potential.
If the condition (\ref{phinecon}) is satisfied, the field evolves toward
smaller values in the region $0<r<r_2=1/m_A$.
In this case the field derivative $\phi'(r)$ needs to change sign 
at $r=r_3$ ($r_2<r_3<r_1$) for the realisation of the thin-shell solution.
If $\phi'(r)$ is positive in the region $0<r<r_1$, the field dynamics 
is similar to the one in the Minkowski background, discussed 
in the previous subsection.

\begin{figure}
\begin{centering}
\includegraphics[height=3.0in,width=3.2in]{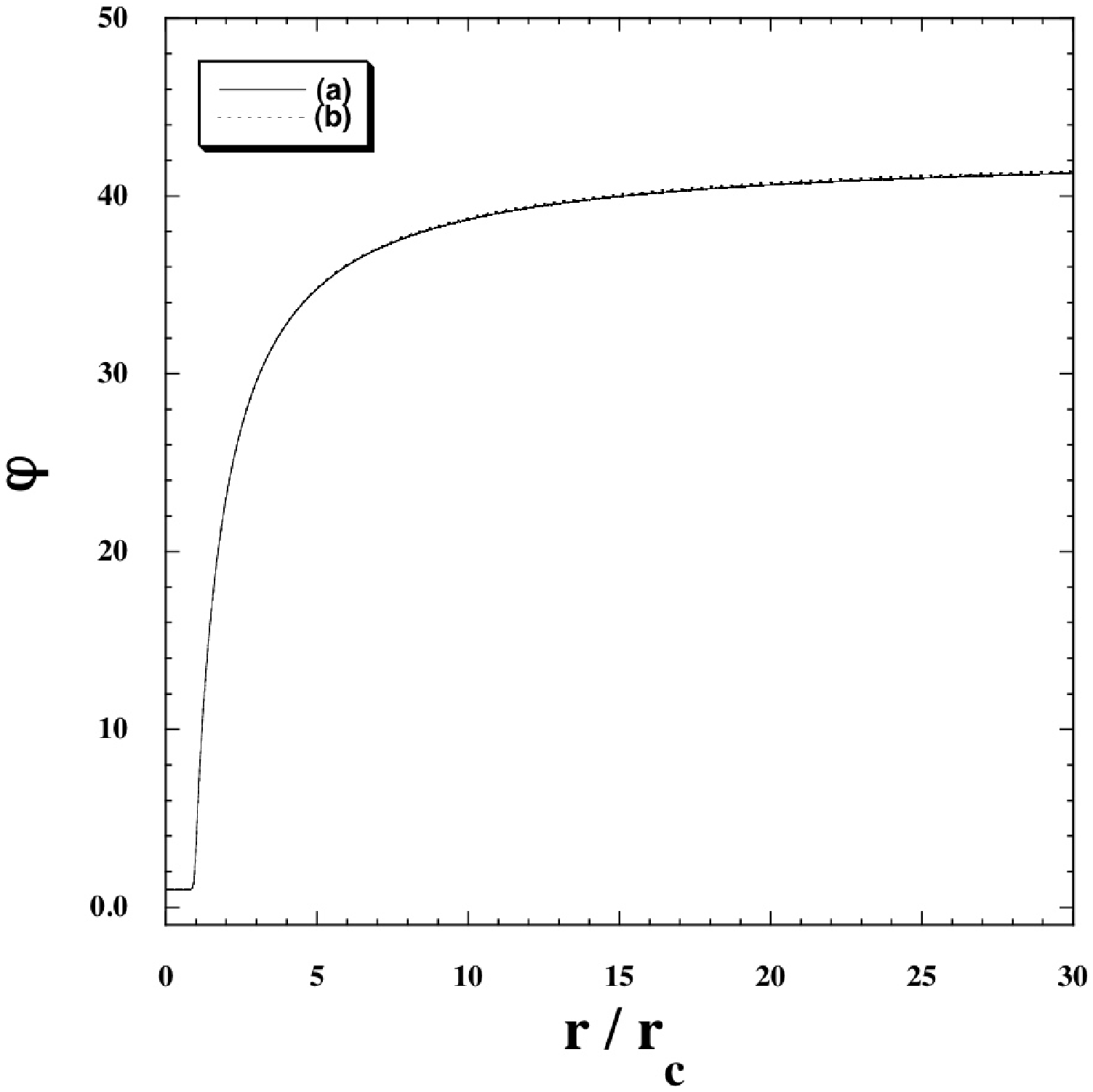}
\includegraphics[height=3.0in,width=3.2in]{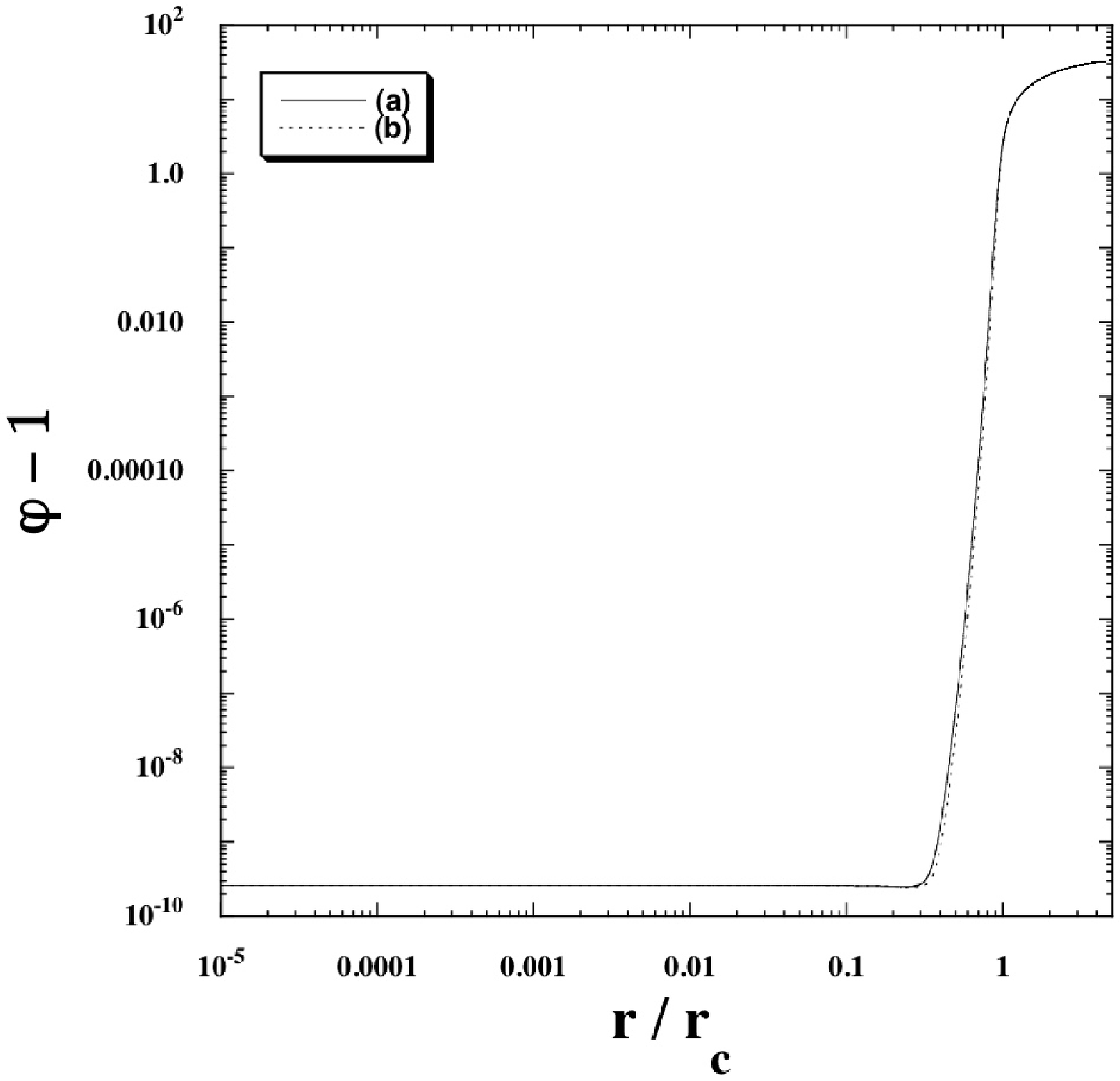}  
\par\end{centering}
\caption{The thin-shell field profile for $\Phi_c=7.0 \times 10^{-10}$,
$n=3$, $Q=1$, $\Delta r_c/r_c=0.085$ and $m_A r_c=40.0$.
This case corresponds to $\rho_A/\rho_B=3.3 \times 10^6$,
$\phi_A=1.05 \times 10^{-11}$, $\phi_B=4.48 \times 10^{-10}$ 
and $\epsilon_{\rm th}=0.104$.
The boundary condition for the field at $x_i=10^{-5}$ is 
$\varphi(x_i)-1=2.615210 \times 10^{-10}$, which is
slightly larger than the analytic value 
$\varphi(x_i)-1= 2.615179 \times 10^{-10}$ 
that comes from Eq.~(\ref{phiini}).
The derivative $\varphi'(x_i)$ is chosen to be 
the same as the analytic value.
The left panel shows $\varphi$ in the region $0<r/r_c<30$, while 
the right panel depicts $(\varphi-1)$ in the region $0<r/r_c<5$
with log scales in the vertical and horizontal axes. 
The black curve (a) shows the numerically integrated solution, while 
the dotted curve (b) is the analytic field profile 
given in Eqs.~(\ref{phiini})-(\ref{phiini3}).
The solution approaches the asymptotic value $\varphi_B=42.705$.}
\label{fig3}
\end{figure}

Let us now consider the case $\phi'(r)<0$
in the region $0<r<r_3$.
As long as $m_A r_1 \gg 1$ this situation naturally appears
even in weak gravity backgrounds such as in the case of
the Earth or the Sun.
In Fig.~\ref{fig3} we present an example of a numerically 
integrated field profile corresponding to the 
gravitational potential of the Earth with 
$\Phi_c=7.0 \times 10^{-10}$ and
$n=3$, $Q=1$, $\Delta r_c/r_c=0.085$ and $m_A r_c=40.0$.
We choose the boundary condition of $\phi$ at $x_i=10^{-5}$
to be slightly larger than the analytic value derived from 
Eq.~(\ref{phiini}), so that the numerical solution
approaches the field value $\varphi_B=\phi_B/\phi_A=42.705$
asymptotically. The reason for this choice is that 
matching two analytic solutions at $r=r_1$
leads to an overestimation of the field value by neglecting 
the term $V_{,\phi}$ relative to the term $Q\rho_A$
in the region $r_1<r<r_c$. 
The resulting field profile is sensitive to a slight change of 
the boundary condition for $\varphi (x_i)$.
This shows the importance to derive analytic solutions for
finding appropriate thin-shell solutions, as we have done
in previous sections.

In the region $0<r<r_3 \simeq 0.26r_1$ the derivative $\phi'(r)$
is negative. It changes sign at $r=r_3$ and the field begins to grow
in the region $r>r_3$ for increasing $r$. 
This behaviour is confirmed in the right panel of Fig.~\ref{fig3}.
Around the surface of the body the field acquires sufficient 
kinetic energy so that it climbs up the potential hill toward
$\phi=\phi_B$. The left panel of Fig.~\ref{fig3} shows that 
the numerical solution outside the body agrees well with 
the analytic thin-shell solution given in Eq.~(\ref{phiini3}).
Thus the chameleon mechanism is present in the 
relativistic background with weak gravity ($\Phi_c \ll 1$).

\begin{figure}
\begin{centering}
\includegraphics[height=3.0in,width=3.2in]{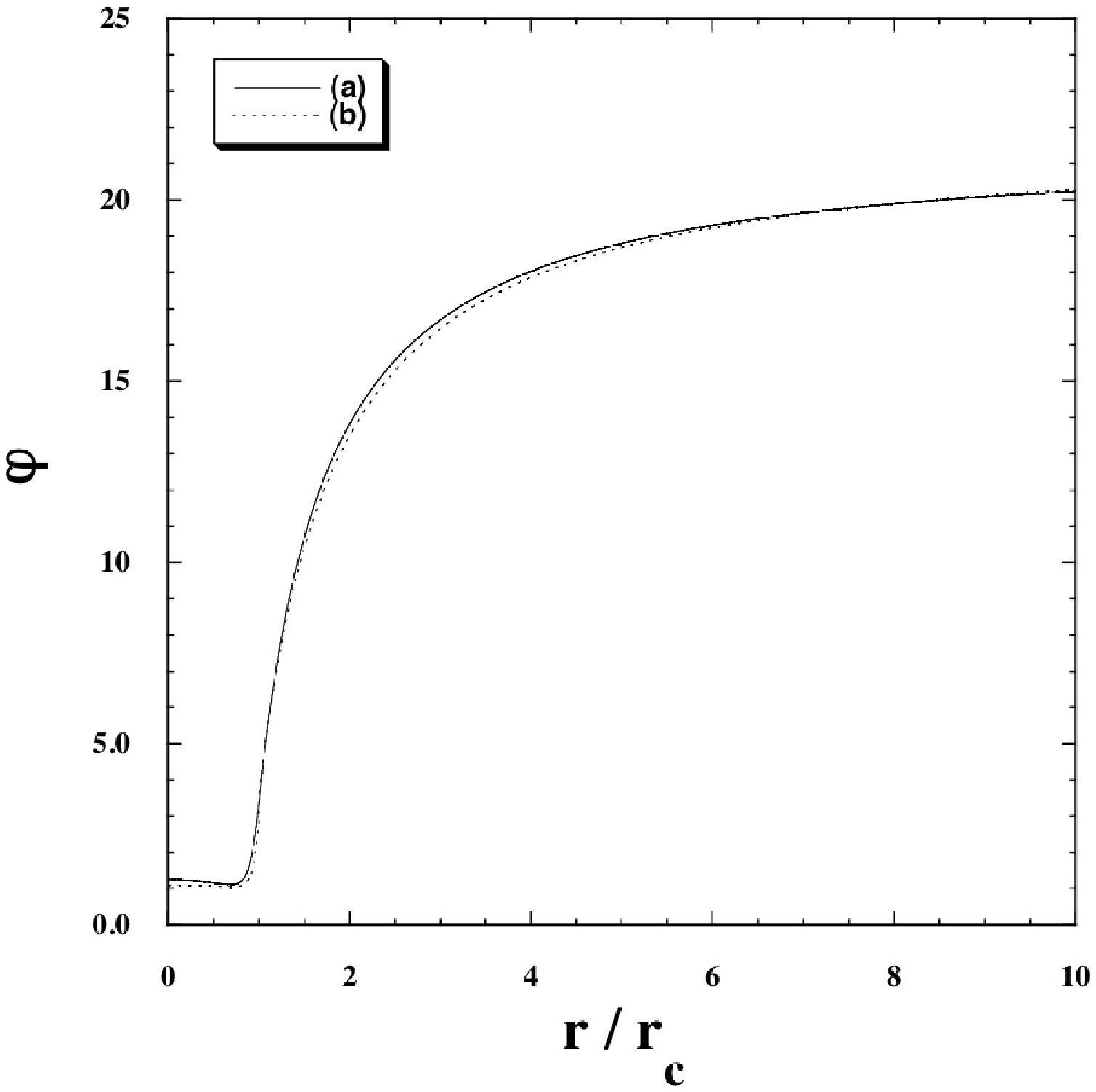}
\includegraphics[height=3.0in,width=3.2in]{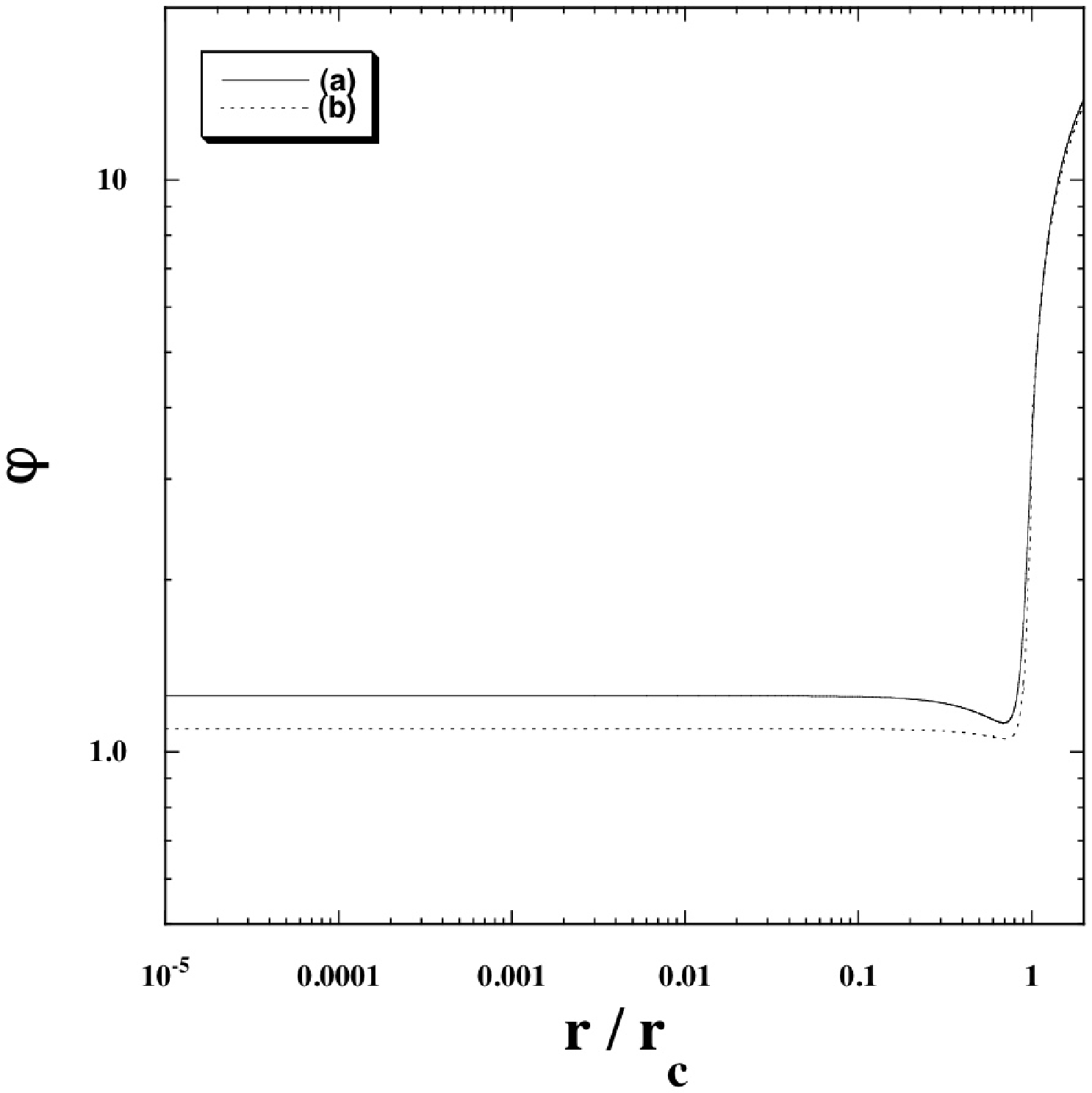}  
\par\end{centering}
\caption{The thin-shell field profile for $\Phi_c=0.2$,
$n=2$, $Q=1$, $\Delta r_c/r_c=0.1$ and $m_A r_c=20.0$.
This case corresponds to $\rho_A/\rho_B=1.04 \times 10^4$,
$\phi_A=8.99 \times 10^{-3}$, $\phi_B=1.97 \times 10^{-1}$ and
$\epsilon_{\rm th}=1.56 \times 10^{-1}$.
The boundary condition of the field at $x_i=10^{-5}$ is 
$\varphi(x_i)=1.2539010$, which is
larger than the the analytic value 
$\varphi(x_i)=1.09850009$ 
that comes from Eq.~(\ref{phiini}).
The derivative $\varphi'(x_i)$ is chosen to be 
the same as the analytic value.
The left panel depicts $\varphi$ in the region $0<r/r_c<10$, while 
the right panel depicts $\varphi$ in the region $0<r/r_c<2$
with log scales in the vertical and horizontal axes. 
The black curve (a) and the dotted curve (b) correspond to 
the numerically integrated solution and 
the analytic field profile (\ref{phiini})-(\ref{phiini3}), respectively.
The numerical solution outside the body recovers the analytic
field profile (\ref{phiini3}).}
\label{fig4}
\end{figure}

For larger gravitational potential $\Phi_c$, the effect of
the relativistic pressure becomes stronger
around the centre of the body.
This leads to the rapid evolution of the field $\phi$ 
toward smaller values.
We also note that the analytic thin-shell solution
(\ref{phiini})-(\ref{phiini3}) begins to lose its accuracy 
in the stronger gravitational backgrounds
with $\Phi_c \gtrsim 0.1$.
If we run our numerical code by choosing boundary 
conditions for $\phi(r)$ and $\phi'(r)$ around the centre of 
the body determined by Eq.~(\ref{phiini}), 
the solutions with $\Phi_c \gtrsim 0.1$ typically keep evolving 
toward smaller $\phi$ regions by overshooting the effective 
potential maximum at $\phi=\phi_A$.
However, if we choose boundary conditions for $\phi$
which are larger than the one given by the corresponding
analytic value, we find that it is possible 
to reproduce the analytic thin-shell solution (\ref{phiini3})
outside the body even for $\Phi_c \sim 0.1$.
The need for the choice of larger $\phi$ partially comes from 
the overestimation of the field around $r=r_1$, 
as was explained above.
Moreover, since the pressure is underestimated in our linear
expansion of $\Phi_c$, we need to choose
values of $\phi$ larger than the corresponding analytic values 
in order to prevent the 
field from entering the region $\phi<\phi_A$.

In Fig.~\ref{fig4} we plot an example of the numerical solution 
for $\Phi_c=0.2$, $n=2$, $Q=1$, $\Delta r_c/r_c=0.1$ and $m_A r_c=20.0$,
together with the corresponding analytic field profile.
We have used the boundary condition 
$\varphi(x_i=10^{-5})=1.2539010$, 
which is larger than the analytic value 
$\varphi(x_i=10^{-5})=1.09850009$ 
estimated by Eq.~(\ref{phiini}).
We note again that the resulting field profile is sensitive to 
the change of boundary conditions.
As can be seen from the right panel of Fig.~\ref{fig4} 
the derivative $\phi'(r)$ is negative in the region $0<r/r_c<0.69$.
The field grows for increasing $r$ in the region $r/r_c>0.69$ 
so that it enters the thin-shell regime for $r/r_c>0.9$.
The left panel of Fig.~\ref{fig4} shows that 
the numerical solution outside the body agrees well 
with the corresponding analytical solution. 
The solution asymptotically 
approaches the field value $\phi_B/\phi_A=21.844$.

We have also carried out numerical simulations for other model 
parameters in the strong gravitational backgrounds.
We find that thin-shell solutions are present for $\Phi_c \lesssim 0.3$, 
which marginally includes the case of neutron stars.
When $\Phi_c \gtrsim 0.3$, however, the field continues to 
evolve toward smaller $\phi$
and overshoots the effective potential maximum at $\phi=\phi_A$ (i.e., $\varphi=1$) 
unless the boundary condition around the centre of the body 
is chosen to be $\phi/\phi_A \gg 1$.
The evolution of the field is typically followed by the rapid roll-down along 
the potential toward the singularity at $\phi=0$ (as in the
numerical simulations of Kobayashi and Maeda \cite{Kobayashi} 
for the $f(R)$ dark energy model of Starobinsky \cite{Star}).
Since the ratio $p_m (r)/\rho_A$ is of the order 
of $\Phi_c$ around the centre of the body,
the pressure force is so strong that the field typically
overshoots the effective potential maximum in such cases.
We stress here that in strong gravitational 
backgrounds with $\Phi_c={\cal O}(1)$ a separate analysis is 
required without recourse to the analytic solutions derived here which are
valid only in the regimes with $\Phi_c \lesssim {\cal O}(0.1)$.

Finally we note that the distance $r_3$ at which $\phi'(r_3)=0$
gets smaller for decreasing $m_A$ (see Table I).
This may suggest that it is possible to avoid the overshooting of the field
by choosing smaller values of $m_A$.
However, the parameter $\Delta r_c/r_c$ needs to satisfy 
the conditions $\Delta r_c/r_c \ll 1$ and $\Delta r_c/r_c \gg 1/m_A r_c$.
This implies that we can not choose the values of $m_A r_c$
that are smaller than the order of 10.
Thus when $\Delta r_c/r_c \lesssim 0.1$ and $m_A r_c \gtrsim 10$,
it is typically difficult to obtain thin-shell solutions
for $\Phi_c \gtrsim 0.3$, whereas thin-shell solutions are present
for $\Phi_c \lesssim 0.3$.

\section{Conclusions}
\label{conclude}

In this paper we have studied the behaviour of the chameleon 
scalar field $\phi$ in the relativistic gravitational background of the 
spherically symmetric space time.
The gravitational potentials $\Phi$ and $\Psi$ are found
analytically under the conditions that the density of the central compact 
object is constant and that the energy density of the chameleon 
field is much smaller than that of the matter.
Using the gravitational potential $\Phi_c$ at the surface of the body
as a linear expansion parameter we have derived
the scalar-field equation (\ref{fieldeq}).

The solutions to the field equations can be obtained 
by considering the perturbation $\delta \phi$ about 
the corresponding solution $\phi_0$ in the Minkowski background.
In the region $0<r<r_1$ the field exists around 
the minimum of the effective potential 
$V_{\rm eff} (\phi)=V(\phi)+Q \rho_A \phi$ inside the body.
The thin-shell case corresponds to settings in which 
$r_1$ is close to the radius $r_c$ of the body.
In the region $r_1<r<r_c$ the coupling term $Q \rho_A \phi$
dominates over the effective potential, which leads to 
rapid changes in the field.
Using linear expansions in terms of $\Phi_c$ and $\delta \phi$
we have derived the solutions of the field equation in the regions
$0<r<r_1$ and $r_1<r<r_c$, see Eqs.~(\ref{phiso1}) 
and (\ref{phiso2}).
Outside the body, the kinetic energy of the field dominates 
over its potential energy, so that the approximate solution
in this region is given by Eq.~(\ref{phiso3}).

We have matched the three solutions at the distances $r_1$ 
and $r_c$ subject to boundary conditions (\ref{boundary})
and have derived the analytical thin-shell field profile 
given by Eqs.~(\ref{phithin1})-(\ref{phithin3}). 
In discussing the analytical field profile, we have 
considered the case $Q>0$ for simplicity. 
Compared to the result of the Minkowski spacetime,
the field $\phi$ around the centre of the body is shifted due
to the presence of a relativistic pressure.
For larger values of $\Phi_c$ and $m_Ar_c$, the field derivative 
$\phi'(r)$ becomes negative in the region $0<r<r_3~(<r_1)$. 
For values of $r>r_3$, $\phi'(r)$ becomes positive
and the field $\phi(r)$ begins to grow with 
increasing $r$. 
As long as the condition $\Delta r_c/r_c \gg 1/(m_A r_c)$
is satisfied, the field acquires sufficient kinetic energy 
in the thin-shell regime in order to climb up the potential 
hill outside the body.

For the class of potentials $V(\phi)=M^{4+n}\phi^{-n}$ we have carried 
out numerical simulations by using the information provided
by the analytic field profile in order to set the boundary conditions 
around the centre of the body.
In the Minkowski background ($\Phi_c=0$) the thin-shell 
field profile outside the body can be recovered numerically 
by choosing the boundary condition of the field to be
larger than the corresponding analytic value.
The reason for this comes from the fact that the analytic solution 
overestimates the field value in the region $r_1<r<r_c$
by neglecting the term $V_{,\phi}$ relative to $Q\rho_A$.

In the relativistic gravitational backgrounds with $\Phi_c \lesssim 0.3$
we have also confirmed the presence of thin-shell solutions 
numerically. While there exists a region in which $\phi'(r)$ is 
negative inside the body, it is possible to realize thin-shell solutions
if the derivative $\phi'(r)$ changes sign at a distance $r=r_3$
smaller than $r_1$. For larger $\Phi_c$ the distance $r_3$ tends
to increase so that the effect of the relativistic pressure is stronger
inside the body. We note that our analysis does not cover the case of
extremely stong gravitational backgrounds with $\Phi_c$ of the order 
of unity. This requires a separate detailed analysis which incorporates
the formation of black holes.

Finally we note that realistic stars have densities $\rho_A(r)$
that globally decrease as a function of $r$.
It would be expected that this decreasing density may work 
as a counter term to the relativistic pressure 
around the centre of the body [see Eq.~(\ref{be1})]. 
It would be of interest to see whether thin-shell 
solutions are present in such realistic cases with 
strong gravitational backgrounds.
We shall return to this question in future.

\section*{ACKNOWLEDGEMENTS}
We thank Justin Khoury, Tsutomu Kobayashi 
and David Langlois for useful correspondences and discussions.
ST thanks financial support for JSPS (No.~30318802).
ST is thankful for kind hospitality during his stay at 
Queen Mary University of London at which this work was initiated.


\end{document}